\documentclass[conference]{IEEEtran}
\IEEEoverridecommandlockouts
\usepackage{cite}
\usepackage{amsmath,amssymb,amsfonts}
\usepackage{algorithmic}
\usepackage{graphicx}
\usepackage{textcomp}
\usepackage{xcolor}
\usepackage[ruled,linesnumbered]{algorithm2e}
\usepackage{pifont}
\usepackage{booktabs}
\usepackage{multirow}
\usepackage{amsthm}
\newtheorem{theorem}{Conclusion}
\usepackage{float}

\def\BibTeX{{\rm B\kern-.05em{\sc i\kern-.025em b}\kern-.08em
    T\kern-.1667em\lower.7ex\hbox{E}\kern-.125emX}}

\begin{document}

\title{A Scale-out Decentralized Blockchain Ledger System for Web3.0\\
}

\author{\IEEEauthorblockN{Lide Xue}
\IEEEauthorblockA{\textit{University of Science and Technology of China} \\
Hefei, China \\
xldxld@mail.ustc.edu.cn}
\and
\IEEEauthorblockN{Wei Yang}
\IEEEauthorblockA{\textit{University of Science and Technology of China} \\
Hefei, China \\
qubit@ustc.edu.cn}
\and
\IEEEauthorblockN{Wei Li}
\IEEEauthorblockA{\textit{University of Science and Technology of China} \\
Hefei, China \\
wei123@mail.ustc.edu.cn}
}

\maketitle

\begin{abstract}
The development of underlying technologies in blockchain mostly revolves around a difficult problem: how to enhance the performance of the system and reduce various costs of nodes (such as communication, storage and verification) without compromising the system's security and decentralization. Various layer-1 and layer-2 protocols have provided excellent solutions for this challenge. However, they cannot yet be considered as a ``silver bullet". This paper proposes EZchain---a novel decentralized ``scale-out" ledger system designed for web3.0, aiming to enable blockchain technology to truly support ledger applications in large-scale fully decentralized networks. Without compromising security and decentralization, EZchain successfully accomplishes the following milestones: 1) Scalability: The theoretical throughput of EZchain can be infinitely expanded, nearly unaffected by bandwidth and other resource constraints. 2) Consumer-Grade Hardware Compatibility: EZchain is designed to be compatible with consumer-grade hardware, supporting storage, computation, and verification requirements. 3) Efficient Transaction Confirmation: EZchain strives to maintain transaction confirmation delays within one minute.
Our prototype experiment demonstrates that under typical daily bandwidth network conditions, EZchain's performance in all aspects approaches that of the accounts in centralized payment systems. This provides a solid infrastructure for realizing mobile payments in web3.0.
\end{abstract}


\section{Introduction}
In blockchain applications, achieving the same central server-level performance of web2.0 without sacrificing decentralization and security has always been a key bottleneck, also known as the blockchain impossible triangle.

\begin{figure}[htp!]
    \centering
    \includegraphics[width=0.95\linewidth]{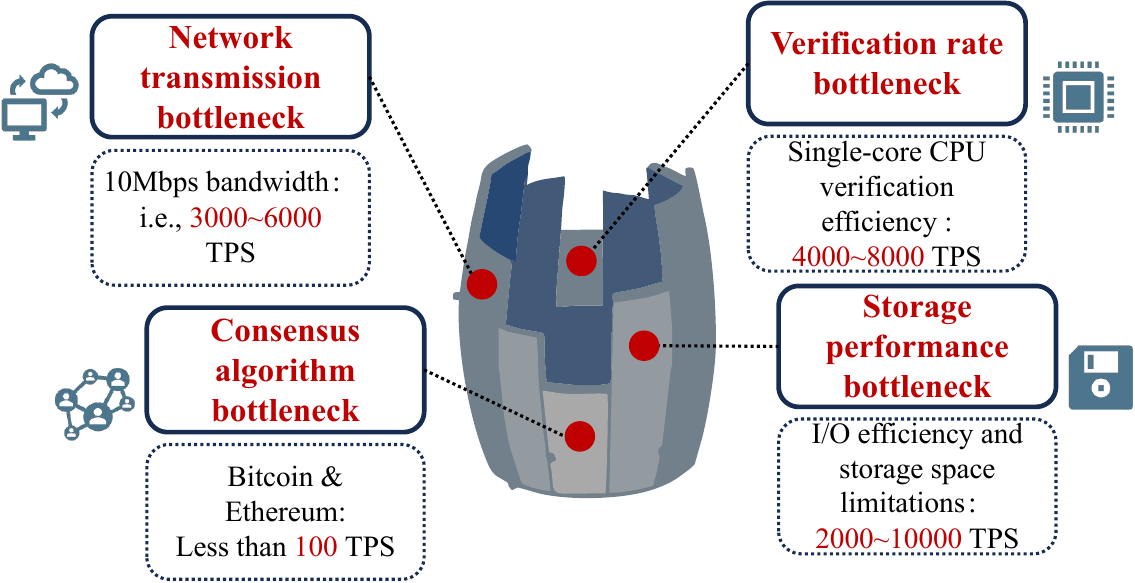}
    \caption{The ``barrel effect" of various bottlenecks faced by blockchain and web3.0 applications.}
    \label{fig:barrel effect}
\end{figure}

Given a blockchain system with $n$ nodes, global consensus requirements inherently involve at least \(O(n)\) message and storage complexity, with some BFT-like algorithms even demanding \(O(n^3)\) in specific scenarios. This complexity induces numerous system bottlenecks,  as depicted in Figure~\ref{fig:barrel effect}. For instance, in distributed ledgers, transaction validation and double-spending checks necessitate tracing the entire blockchain history for confirmation\footnote{An alternative involves utilizing a ``world state" snapshot for verification; however, this still depends on the validity confirmation of the``world state", either self-validated or by other nodes. In a truly decentralized context,``world state" validation also entails tracing the complete blockchain history.}. Furthermore, incorporating a transaction into a new block requires broadcasting it to all consensus nodes for validation. From a storage perspective, each full-consensus node essentially mirrors a central server, redundantly storing the world state and transaction history backups. To overcome these challenges, various approaches like sharding, off-chain, and cross-chain solutions have been explored, often trading off a degree of security and decentralization for enhanced system efficiency. Addressing the impossible triangle and these multifaceted bottlenecks is crucial for developing a robust infrastructure for the future web3.0.

Concerning the above challenges, in this paper we introduce EZchain, a blockchain solution characterized by its high-performance, decentralization, and robust security. Distinct from pseudo-decentralization, layer-2, and off-chain concepts, EZchain adheres strictly to the principles of decentralization, achieving layer-1 performance breakthroughs. EZchain theoretically offers:
\begin{enumerate}
\item A constant block size of approximately 0.5 Mb, capable of accommodating an unlimited number of transactions.
\item The performance of ``scale-out" exhibits unbounded system throughput.
\item Transaction confirmation within seconds.
\item Feasibility for consumer-level storage, messaging, and verification costs for both consensus and account nodes.
\item Uncompromised security and decentralization.
\end{enumerate}

EZchain implements a pioneering consensus algorithm designed to minimize consensus, transmission, and storage information. It introduces a novel data structure, ``value", distinct from UTXO model in Bitcoin~\cite{BitcoinWiki2019} or account balance model. This unique combination of data structure and algorithm, to our knowledge, represents an unprecedented approach in the blockchain scalability field. The underlying rationale for EZchain's remarkable performance lies in the relative stability of the ``value" quantity, as opposed to an ever-increasing transaction volume. This shift in focus to value transfer offers unique advantages and optimization strategies over traditional methods. To empirically assess EZchain's effectiveness, we developed a prototype simulation system\footnote{github.com/Re20Cboy/Ezchain-py}, whose experimental results corroborate EZchain's claims regarding throughput, scalability, transaction confirmation speed, and storage efficiency for nodes.

\section{Related works}
\label{sec: Related works}

Among works closely related to EZchain, the Vapor blockchain, introduced by Ren et al.~\cite{VAPOR2018,SpontaneousSharding}, stands out. Vapor pioneered the ``values"-like concept, termed ``adaptive sharding". Despite its claimed scale-out performance, our approach diverges in key aspects. Firstly, Vapor's block size varies and escalates with node count expansion, as detailed in its block data structure design~\cite{SpontaneousSharding}. Secondly, lacking targeted algorithmic refinement for value transfer, Vapor does not match EZchain's performance levels. More importantly, as the system operates, the communication and storage complexity of Vapor will tend towards $O(n)$, whereas EZchain will approach a constant independent of $n$.


In blockchain scalability research, initial efforts concentrated on expanding block size, enhancing block generation rates, and expediting transaction confirmations in layer-1~\cite{Eyal2016BitcoinNG,Kokoriskogias2016Byzcoin,Algorand,GHOST,HotStuff,bagaria2019prism}. These endeavors primarily optimized Bitcoin's Nakamoto consensus~\cite{Nakamoto2008Bitcoin}. Strategies included decoupling leader election from block consensus~\cite{Eyal2016BitcoinNG} and integrating BFT-like algorithms to reformulate consensus mechanisms~\cite{Kokoriskogias2016Byzcoin,Algorand,HotStuff}. Experimental findings indicate these methods not only achieved breakthroughs in consensus efficiency but also reached transaction processing speeds rivaling centralized systems like PayPal~\cite{Kokoriskogias2016Byzcoin}.

Further advancements in blockchain performance have led to diverse solutions, including sharding, DAG-based blockchains, off-chain mechanisms, cross-chain interoperability, and Zero-Knowledge Proofs technology~\cite{OmniLedger,OHIE,RapidChain,Sharding2016,Chainspace,IOTA,conflux,PHANTOM,LN,Plasma,OptimisticRollup,dilley2016strong,wood2016polkadot,thomas2015protocol,garoffolo2020zendoo,sasson2014zerocash,ben2014succinct,bowe2019recursive,gluchowski2019zk}. Each aims to surmount the inherent bottlenecks in transmission, storage, and verification processes.


Sharding, as a prominent layer-1 innovation, forms the backbone of numerous scalable blockchain systems~\cite{OmniLedger,OHIE,RapidChain,Sharding2016,Chainspace,hong2021pyramid,crain2021red}. This strategy, based on the ``divide and conquer" principle, divides networks, transactions, or states into multiple subsets (shards), enabling consensus algorithms to operate more efficiently across different shards. This approach significantly reduces bandwidth, storage, and computational demands, facilitating enhanced performance. Nonetheless, sharding faces several challenges: firstly, an abundance of cross-shard transactions can adversely affect system efficiency; secondly, conventional consensus security assumptions are not entirely applicable in sharded environments, necessitating additional security models (e.g., OmniLedger, which tolerates up to \(n/4\) Byzantine nodes, assuring shard security with a probability of \(1-10^6\)~\cite{OmniLedger}); furthermore, the intricacy of network sharding algorithms also incurs additional bandwidth and computational overhead.


DAG-based blockchains, diverging from conventional chain structures, utilize a directed acyclic graph (DAG) for block storage~\cite{conflux,PHANTOM,silvano2020iota,tairi20212}. This architecture allows nodes to add blocks efficiently and concurrently, significantly enhancing block generation efficiency and, in theory, approaching the network's transmission capacity. However, two critical challenges persist: 1) Optimizing bandwidth utilization in DAG-based systems demands a delicate balance among consensus efficiency, verification speed, and transmission capability; 2) There is yet to be a universally accepted solution for transaction ordering in DAG-based blockchains.


Off-chain technologies alleviate the main chain's workload, thereby enhancing scalability, efficiency, and privacy within the blockchain system~\cite{LN,Plasma,OptimisticRollup,gavzi2019proof}. Prominent off-chain solutions encompass sidechains, state channels, and lightning networks. Their point-to-point transmission and verification, unencumbered by consensus protocols, present an efficient strategy to overcome various system bottlenecks. Nonetheless, off-chain methods grapple with challenges such as security and trust in environments independent of the main chain's security guarantees, and the reconciliation of off-chain and on-chain data consistency.


Cross-chain technology facilitates interoperability between distinct blockchain networks through intermediate layers or protocols~\cite{dilley2016strong,wood2016polkadot,thomas2015protocol,garoffolo2020zendoo,neu2021ebb,tian2021enabling,sober2021voting}. This technology enables blockchains to share transactions and states. Nevertheless, prevalent cross-chain solutions often depend on centralized validators, a deviation from the decentralization ethos and a compromise on blockchain's inherent security. Furthermore, trustless cross-chain models are associated with elevated costs.

Zero-Knowledge Proofs (ZKPs) are heralded as potential universal solutions for enhancing blockchain scalability and facilitating cross-chain interoperability~\cite{bunz2018bulletproofs,garoffolo2020zendoo,sasson2014zerocash,ben2014succinct,bowe2019recursive,gluchowski2019zk,saleh2021blockchain,grassi2021poseidon,yang2020zero}. They offer the dual benefits of safeguarding user privacy and reducing verification and storage demands. However, the generation and validation of ZKPs, particularly complex ones, can be computationally intensive and time-consuming, posing potential performance bottlenecks for the system.

\section{System Models}
\label{sec: system models}
Following the convention, it is necessary to standardize and define the models of network and threats that the EZchain system is confronted with.

\subsection{Network Model}
\label{subsec: Network Model}
The communication network of EZchain is composed of several consensus nodes and account nodes. All nodes can use broadcasting (e.g., gossip protocol) or peer-to-peer (P2P) mode for message transmission. We assume that this network is weakly synchronous, and communication between non-byzantine fault nodes (i.e., honest nodes) is synchronous. Each honest node has a certain number of honest neighboring nodes, and the network connections between them are robust and unimpeded.

\subsection{Threat Model}
\label{subsec: Threat Model}
EZchain tolerates up to $f = \lambda n$ byzantine nodes, where $\lambda = 1/3$ (for BFT-like backbone consensus algorithm) or $\lambda = 1/2$ (for PoW-like backbone consensus algorithm). The byzantine node can make any deviation from the protocol. In addition, the adversary cannot break our cryptographic assumptions in polynomial-time.

\section{Design of EZchain}
\label{sec:Design of EZchain}

\subsection{Core ideas of EZchain}
\label{subsec:Core ideas of EZchain}

This subsection aims to provide an intuitive and informal introduction to the fundamental concepts underlying the design of EZchain, enabling readers to gain a comprehensive overview of EZchain's system and algorithm design. We shall commence with a story. 

\begin{figure}[htp!]
    \centering
    \includegraphics[width=1\linewidth]{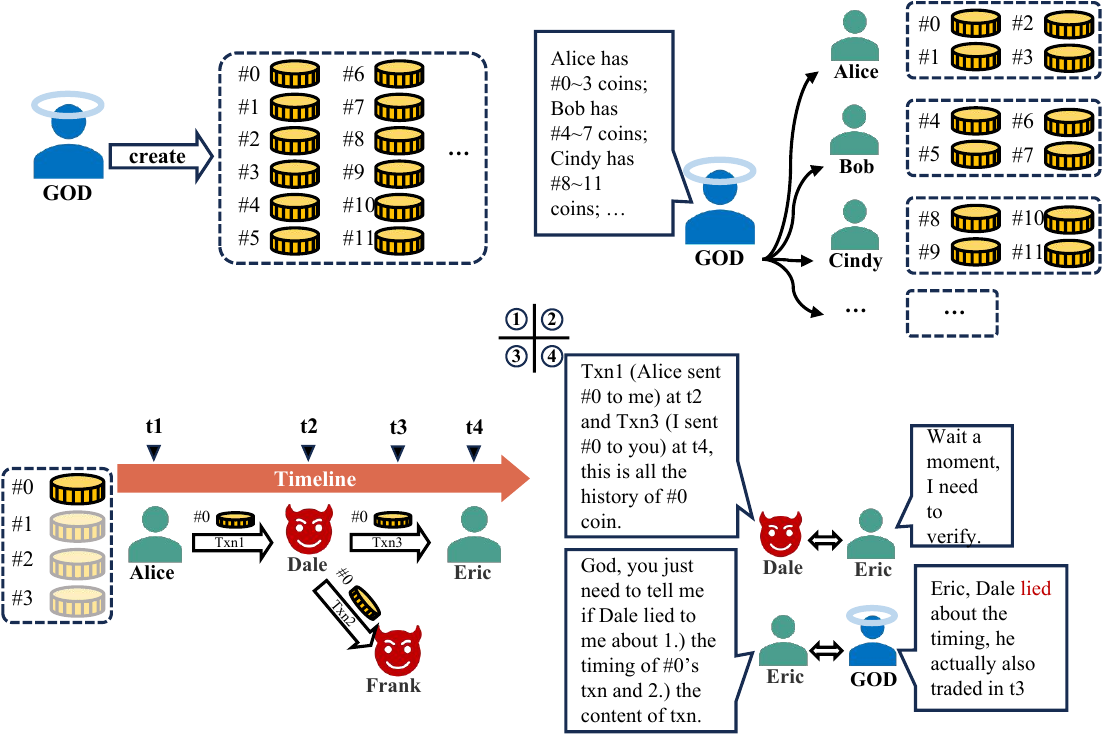}
    \caption{Mythical story about EZchain (1).}
    \label{fig: mythical story of EZchain 1}
\end{figure}

\begin{figure}[htp!]
    \centering
    \includegraphics[width=1\linewidth]{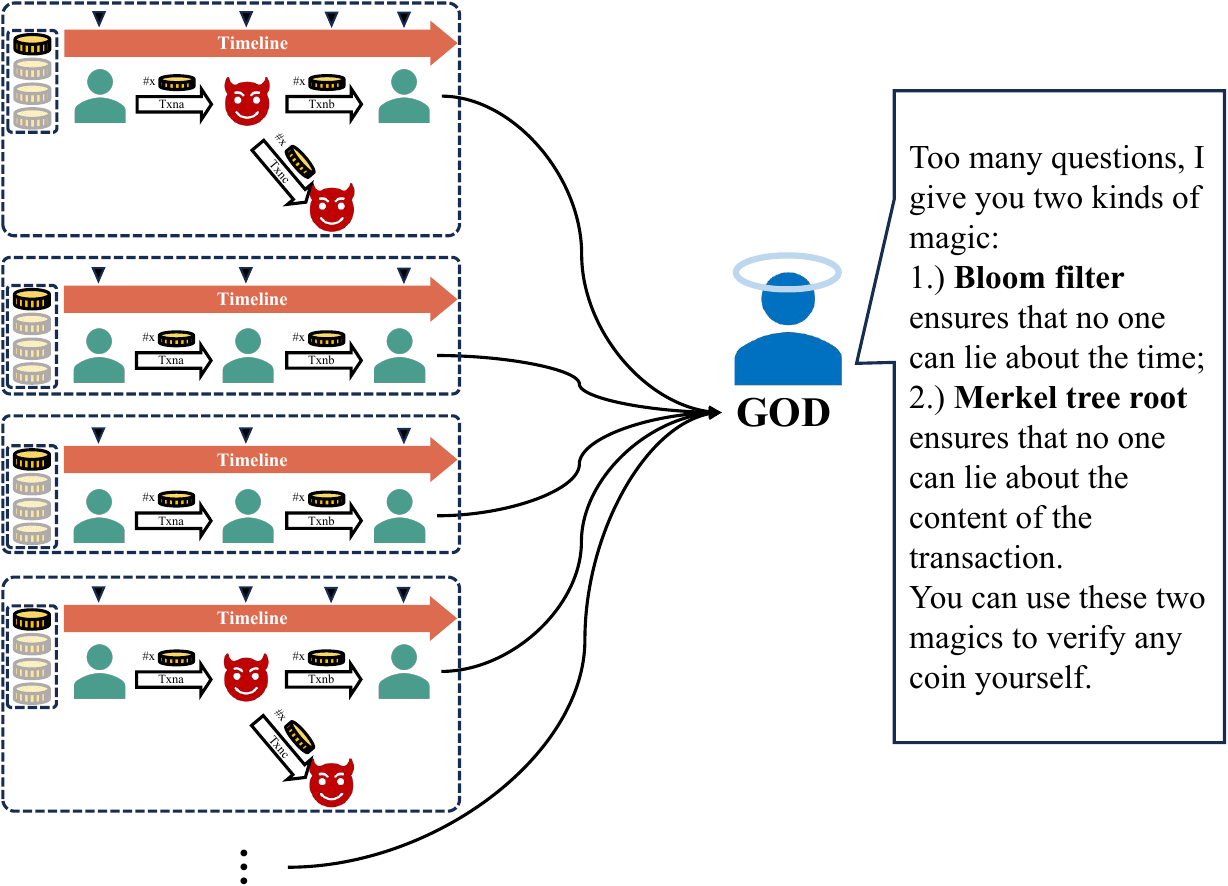}
    \caption{Mythical story about EZchain (2).}
    \label{fig: mythical story of EZchain 2}
\end{figure}

God created a series of coins during the creation and numbered them (Figure~\ref{fig: mythical story of EZchain 1}-\ding{172}). These coins were then distributed among various individuals (Figure~\ref{fig: mythical story of EZchain 1}-\ding{173}). In a specific transaction, Alice used $\# 0$ coin and transferred it to Dale. Upon receiving $\# 0$ coin, Dale covertly passed it to his accomplice, Frank, and subsequently issued an ``empty check" to Eric, falsely claiming that he would transfer $\# 0$ coin to Eric (Figure~\ref{fig: mythical story of EZchain 1}-\ding{174}). Suspicious of Dale's actions, Eric sought clarification from God. God informed Eric that Dale had already spent $\# 0$ coin at time $t_3$. Ultimately, Eric exposed Dale's ``double-spending" scheme (Figure~\ref{fig: mythical story of EZchain 1}-\ding{175}). As more individuals started questioning the authenticity of the received coins, God employed ``two magics" to resolve all the issues (Figure~\ref{fig: mythical story of EZchain 2}).

The core ideas of EZchain are succinctly explained in the previous story:
\begin{enumerate}
	\item Create a new data structure for each value (i.e., coin).
	\item The main chain only needs to validate, consensus, and store these ``two magics"---Merkle tree root and Bloom filter (along with other necessary information).
	\item Recipient can independently verify any transactions through the main chain and the proof provided by the sender.
\end{enumerate}
While the approach of EZchain shares similarities with off-chain and zero-knowledge proofs, its essence is different. Firstly, unlike off-chain solutions, any transaction generated within the EZchain system will be included in a Merkle tree root, ensuring the security of all transactions through the main chain. Secondly, unlike zero-knowledge proof blockchains, EZchain does not rely on any prior zero-knowledge algorithms. Its verification algorithm is concise, allowing for operations such as proof generation and validation to be conducted within a 10-millisecond timeframe while also ensuring that the size of the proof converges to a constant value.

\begin{figure}[htp!]
    \centering
    \includegraphics[width=1\linewidth]{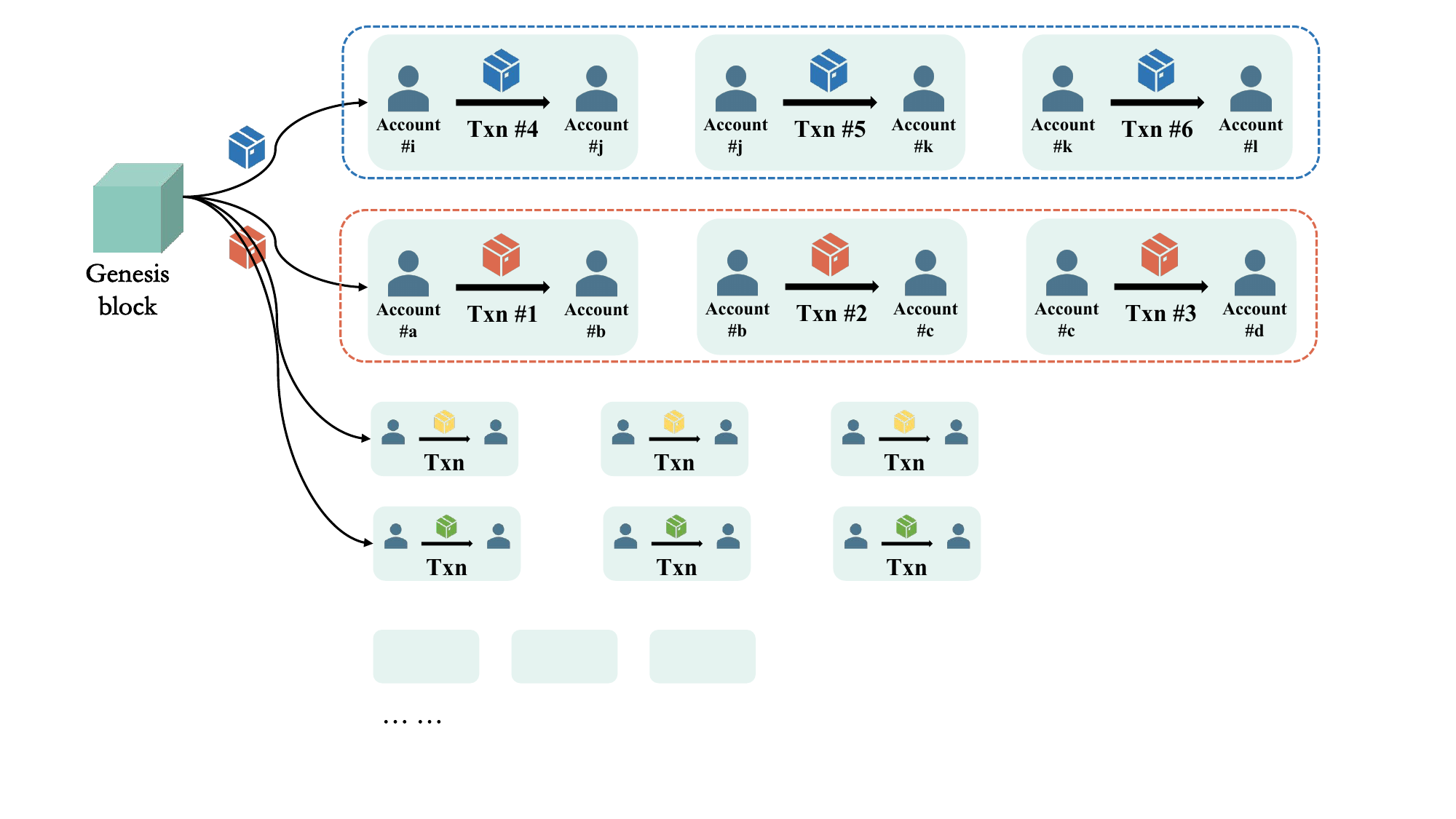}
    \caption{Verification and logical relationship between each transaction.}
    \label{fig:txn ver logical}
\end{figure}

From an information theory perspective, EZchain enables the validation of transaction legitimacy in decentralized networks with malicious nodes without the need for transmitting and storing the entire global history. We have also discovered that a relatively small amount of information (in terms of transmission and storage) can fully support transaction validation in a decentralized and trustless environment\footnote{However, we acknowledge that the current design of EZchain may still require more information than the minimum necessary amount.}. As shown in Figure~\ref{fig:txn ver logical}, a specific value (depicted as a blue cargo box) is continually exchanged among different transactions ($Txn \#4-6$). By receiving this value, account $l$ can verify the legitimacy of $Txn \#6$ by checking the legitimacy of this value. This verification involves checking if all the transactions encountered by this value prior to $Txn \#6$ are legitimate and whether $Txn \#6$ itself is legitimate. Importantly, these verifications are independent of $Txn \#1-3$. Therefore, when verifying $Txn \#6$, the information contained in $Txn \#1-3$ does not need to be transmitted or stored. This concept forms the fundamental idea behind EZchain, similar to the principle of ``Occam's Razor" in blockchain consensus. The goal of EZchain is to eliminate unnecessary information that is irrelevant to consensus and verification, thereby achieving optimized transmission, storage, and verification processes.

\subsection{Overview of EZchain System}
\label{subsec:Overview of EZchain System}

\begin{figure}[h]
    \centering
    \includegraphics[width=1\linewidth]{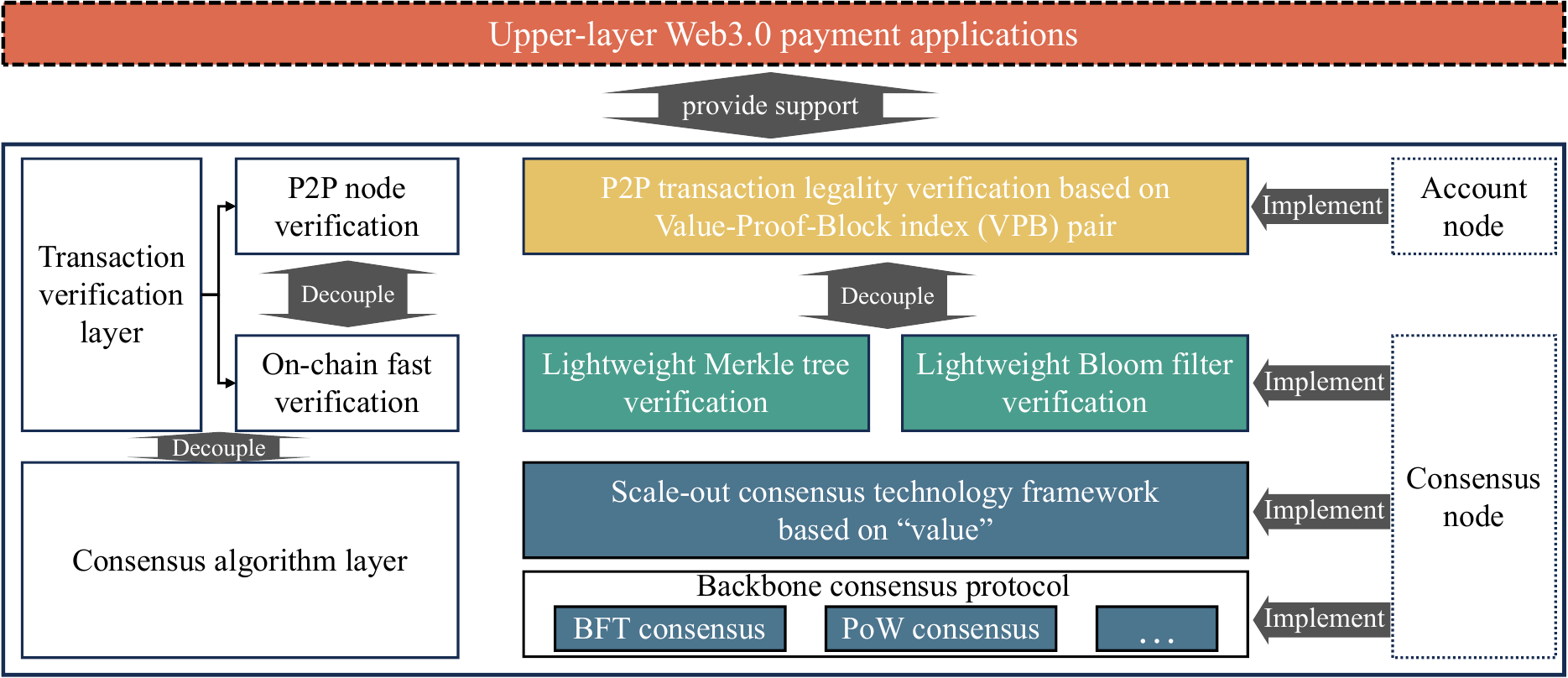}
    \caption{System framework of EZchain.}
    \label{fig:EZchain system framework}
\end{figure}

Overall research ideas and design framework of EZchain are illustrated in Figure~\ref{fig:EZchain system framework}. EZchain's consensus algorithm supports various consensus mechanisms, including Proof of Work (PoW), Byzantine Fault-Tolerant (BFT), Proof of Stake (PoS), and Delegated Proof of Stake (DPoS). In terms of consensus information and data structures, EZchain adopts an innovative ``value"-based mechanism instead of relying on the account or UTXO mechanism. This mechanism primarily focuses on recording and verifying the entire ledger from the perspective of value transfer. On-chain, EZchain utilizes an extremely lightweight data structure wherein each block occupies approximately 0.5 Mb. However, in theory, it can accommodate an infinite number of transactions, achieving ``scale-out" in terms of information stored within a block. Furthermore, the validation of on-chain information is convenient and efficient, requiring only necessary signature and hash validation without the need to backtrack and validate transaction history and logic. This approach also addresses the initial trust issue that arises when new nodes join the EZchain system. Specific transaction validation is deferred to the ``p2p transaction validation" part of account nodes. This means that transaction participants are responsible for verifying the legitimacy of their own transactions, which serves as a positive incentive and decentralizes the validation pressure on consensus nodes.

\begin{algorithm}[htp!]
  \SetAlgoLined
  \KwData{None}
  \KwResult{None}
Miner:
\begin{enumerate}
    \item The miner collects a set of packaged transactions, $TxnPool = \{AccTxn_1, AccTxn_2, ... \}$, from the transaction pool. Here $AccTxn_i = (Sender_i, HASH(Txns_i), SigInfo_i)$, $HASH()$ represents a hash function, $Txns_i$ is a collection of transactions submitted by the account node $Sender_i$, and $SigInfo_i$ denotes the digital signature of $Sender_i$ on $HASH(Txns_i)$.
    \item The miner creates a new block, $Block = (Mtree~Root, Bloom~Filter, Pre~Hash,$ $Time, Miner~Sig, Nonce, Index)$. The $Mtree~Root$ signifies the root of the Merkle tree formed by all $HASH(Txns_i)$ present in the $TxnPool$. The $Bloom~Filter$ comprises the address information of all senders present in the $TxnPool$. Additionally, $Pre~Hash, Time, Miner~Sig, Nonce,$ and $Index$ represent the previous block's hash, timestamp, miner's signature, random number and block's index, respectively.
    \item If the miner successfully wins in the ``mining competition", it broadcasts two messages: $Msg_1 = Block$ and $Msg_2 = SigInfos$. Here, $SigInfos = \{SigInfo_i~|~\forall SigInfo_i \in TxnPool\}$. The miner should prioritize broadcasting $Msg_1$ before $Msg_2$.
    \item Other miners validate the received $Msg_1$ and $Msg_2$ through the following steps: i) check the correctness of all digital signatures; ii) verify that the $Bloom Filter$ matches the $SigInfos$ provided; iii) ensure the ability to reconstruct $Mtree Root$ using the provided $SigInfos$; iv) confirm the absence of duplicate signatures in $SigInfos$ (i.e., a single sender signing two $SigInfo$ within $SigInfos$); v) validate other information including $Pre~Hash, Time, Miner~Sig,$ $Nonce, Index,$ and so on.
\end{enumerate}
  \caption{EZchain main algorithm for miner (PoW version)}
  \label{alg: EZchain miner}
\end{algorithm}

\begin{algorithm}[htp!]
  \SetAlgoLined
  \KwData{None}
  \KwResult{None}
Account:
\begin{enumerate}
    \item When account $i$ initiates a transaction to another account $j$, it creates the transaction $Txn = (Sender, Recipient, Values, Time, SigInfo)$. Here, $Sender$ and $Recipient$ represent the addresses of the sender ($i$) and recipient ($j$), $Values$ is the set of values chosen by the $Sender$ for this transaction, and $Time$ and $SigInfo$ denote the transaction timestamp and the $Sender$'s signature information.
    \item Account $i$ gathers all transactions within a specific time frame and packages them as $Txns_i$. Subsequently, $i$ submits $AccTxn_i = (Sender_i, HASH(Txns_i), SigInfo_i)$ to the transaction pool, awaiting inclusion in a block by a miner.
    \item If the $Block$ is successfully appended to the blockchain, $i$ requests the $Mtree~Proof = \{MTree~Node~|~proving$ $the~presence~of~HASH(Txns_i)~in~the~Merkle~tree\}$ from the miner.
    \item Account $i$ provides $j$ with all values contained in $Txn$'s $Values$ along with the corresponding VPB pairs (refer to Subsection~\ref{subsec:EZchain's data structure} for detail). $j$ then verifies these values by referencing the main chain and the VPB pairs.
    \item Upon successful verification, $j$ accepts the transferred values (i.e., $Values$) from $i$ and stores the relevant proof locally for future use.
\end{enumerate}
  \caption{EZchain main algorithm for account (PoW version)}
  \label{alg: EZchain account}
\end{algorithm}

The overall design logic of EZchain's consensus mechanism is illustrated in Algorithms~\ref{alg: EZchain miner} and~\ref{alg: EZchain account}, respectively. It can be observed that EZchain does not rely on the underlying design of leader election and block generation logic. Therefore, it can adapt to various existing consensus algorithms. In fact, we highly recommend Algorand's consensus mechanism---VRF + BFT~\cite{Algorand}, due to its ability to minimize forks, which aligns well with EZchain's high-speed block generation performance.\footnote{Note that the main algorithm of EZchain does not use the Algorand consensus mechanism for explanation, because the Algorand consensus mechanism is more complex than PoW. Considering readability, we mainly use the PoW version of EZchain for display and explanation.} The specific details of the data structures involved in Algorithms~\ref{alg: EZchain miner} and~\ref{alg: EZchain account} will be explained in the subsequent subsection.

\subsection{EZchain's data structure}
\label{subsec:EZchain's data structure}
Compared with traditional blockchains, EZchain undergoes a meticulous redesign of the data structure regarding blocks, verification, and proofs, as shown in Figure~\ref{fig:value data structure}. The specific explanation is as follows:
\begin{enumerate}
    \item The previously mentioned ``value" is not the same as the token or UTXO in classical blockchain. Its data structure is defined as $value=(Begin~Index, End~Index)$, and it possesses the following characteristics: i) the value is essentially an integer set: $\{x, x \in [Begin~Index, End~Index]~and~x~is~an~integer\}$, ii) different values do not intersect, i.e., $\forall{x}~and~\forall{y},~x \cap y = \varnothing$, iii) the value can be split into smaller sets, and the union of these subsets still equals the original set, iv) the number of values can be calculated based on the $Begin~Index$ and $End~Index$ (i.e., Value's number $= End~Index - Begin~Index + 1$).
    \begin{figure}[htp!]
        \centering
        \includegraphics[width=0.8\linewidth]{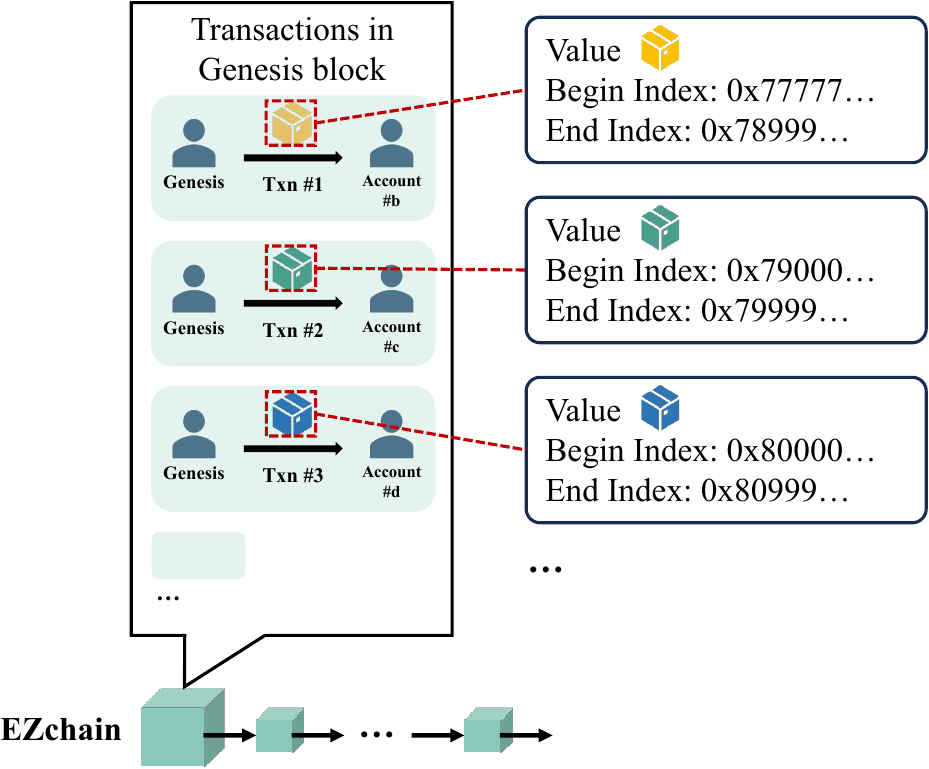}
        \caption{EZchain's value data structure.}
        \label{fig:value data structure}
    \end{figure}

    \item Regarding the data submitted to the transaction pool by an account, which is defined as $AccTxn_i = (Sender_i, HASH(Txns_i), SigInfo_i)$. As shown in Figure~\ref{fig:account node submit data}, account $a$ packages its own initiated transactions, currently pending submission, into a package called $Txns_a$. Subsequently, it computes the hash digest of $Txns_a$ using the function $HASH(Txns_a)$. Next, $a$ signs $HASH(Txns_a)$ to generate the signature, denoted as $SigInfo_a$. Finally, by combining the sender's address information, $Sender_a$, the account transactions are assembled as $AccTxn_a = (Sender_a, HASH(Txns_a), SigInfo_a)$, and they are then submitted to the transaction pool.
    \begin{figure}[htp!]
        \centering
        \includegraphics[width=0.95\linewidth]{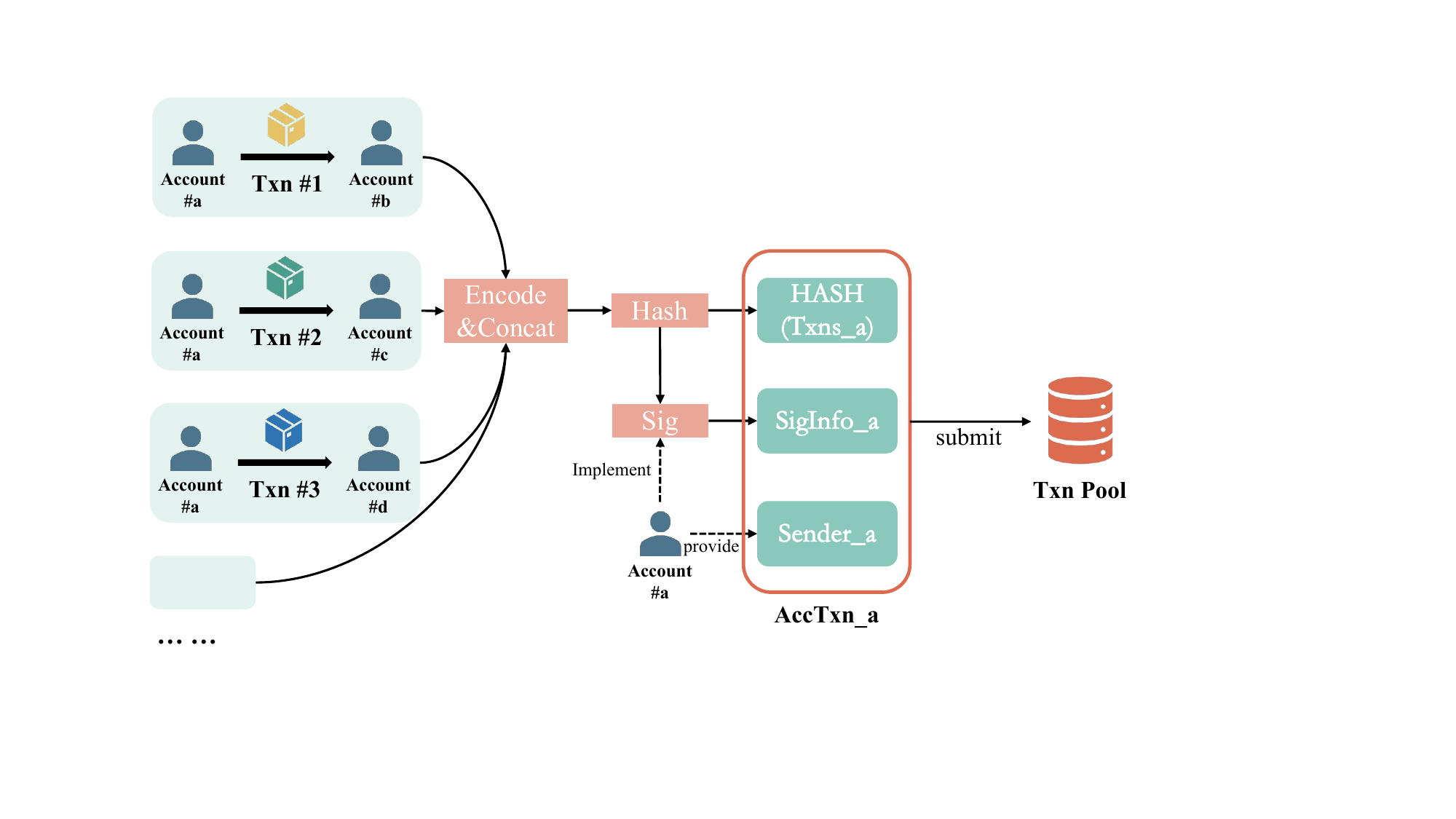}
        \caption{Data structure submitted by EZchain account nodes to the transaction pool.}
        \label{fig:account node submit data}
    \end{figure}

    \item In terms of the block structure, the EZchain's block contains the following information: $Block = (Mtree~Root, Bloom~Filter, Pre~Hash, Time,$ $ Miner~Sig, Nonce, Index)$ (as shown in Figure~\ref{fig:Merkle tree and Bloom filter}). These components represent the merkle tree root, bloom filter, previous block's hash, timestamp, miner's signature, nonce and block's index, respectively. Now let us provide a detailed explanation of two unique data structures: $Mtree~Root$ and $Bloom~Filter$: i)  $Mtree~Root$: the $Mtree~Root$ is obtained by collecting all the $HASH(Txns_i)$s from the transaction pool. These $HASH(Txns_i)$s are the leaf nodes used to construct a merkle tree. The root node of this tree represents the $Mtree~Root$ stored within the block. ii) $Bloom~Filter$: the $Bloom~Filter$ is a data structure that includes the address information of all senders within the $AccTxn$ collection. Other data structures within the block are similar to those found in classical blockchain systems.
    \begin{figure}[htp!]
        \centering
        \includegraphics[width=0.95\linewidth]{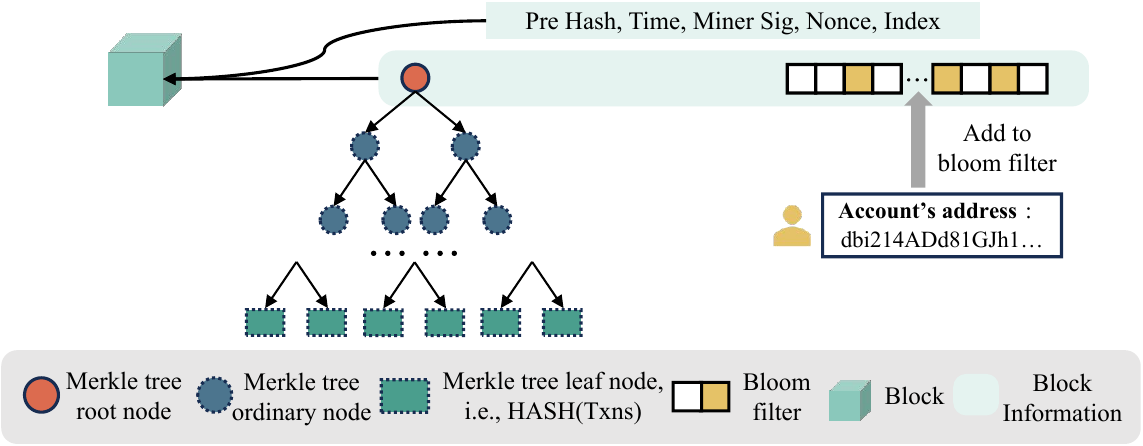}
        \caption{Merkle tree root and Bloom filter in EZchain's block.}
        \label{fig:Merkle tree and Bloom filter}
    \end{figure}

    \item In terms of data broadcast by the miner, which are $Msg_1=Block$ and $Msg_2=SigInfos$, $Block$ has been described in Figure~\ref{fig:Merkle tree and Bloom filter}, while $SigInfos$ is a set comprising all $SigInfo_i$ in $TxnPool$, where $TxnPool$ is defined as the set of ${AccTxn_1, AccTxn_2, …}$. Figure~\ref{fig:Miner broadcast data} provides a visual representation of the data structure and content of the two messages broadcasted by the Miner.
    \begin{figure}[htp!]
        \centering
        \includegraphics[width=1\linewidth]{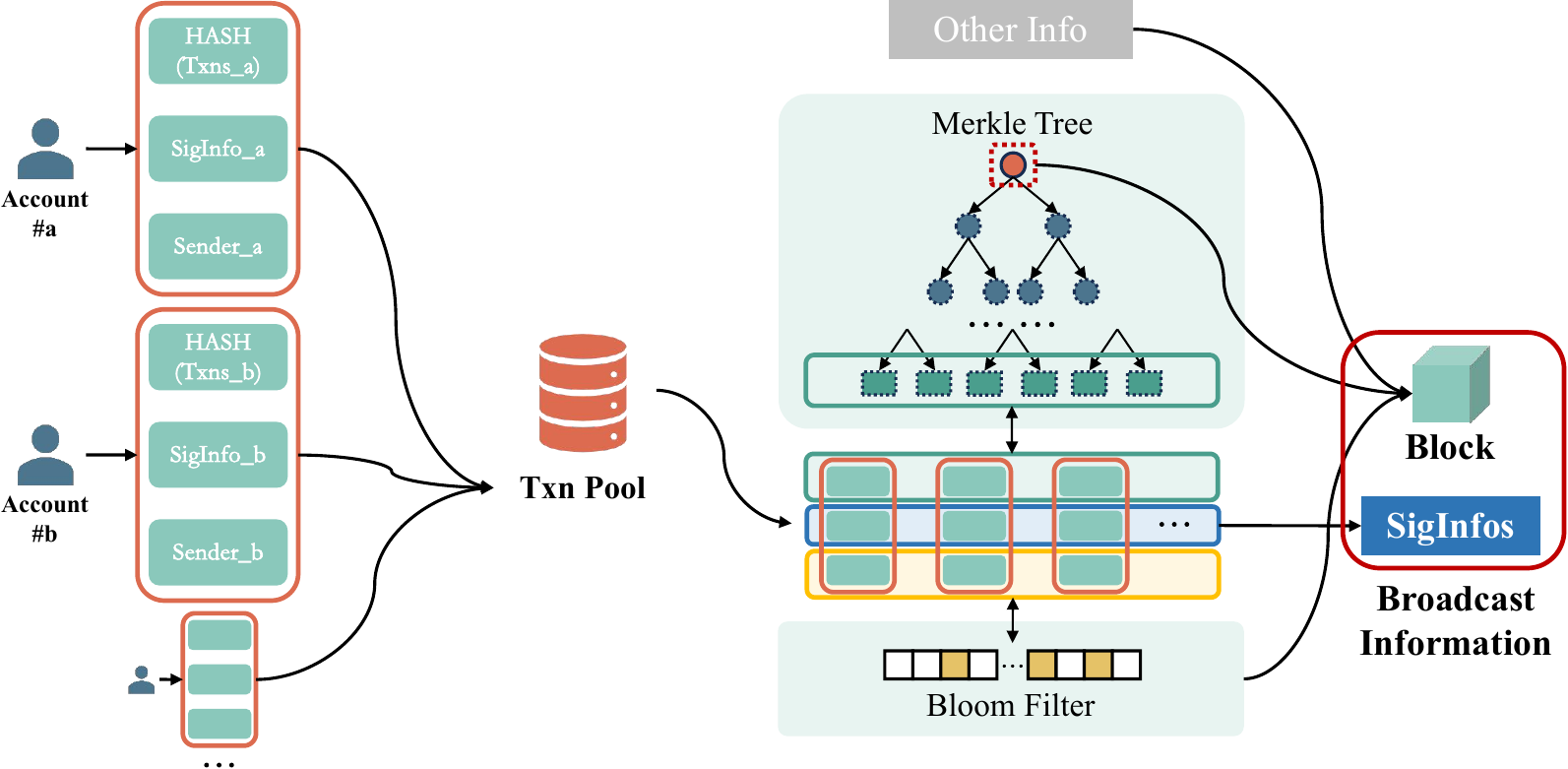}
        \caption{Data structure of Miner broadcast information.}
        \label{fig:Miner broadcast data}
    \end{figure}
    \item In terms of data presented by $Account_i$ to the recipient during transactions, EZchain utilizes a data structure known as the ``VPB pair (Value-Proof-Block Index pair)". In this structure, the term ``Value" represents the payment value chosen by $Account_i$ for the transaction. ``Proof" encompasses the following elements: i) $Mtree~Proof$: it comprises a list of tree nodes in the Merkle Tree ($MTree$) that provide evidence for the existence of $HASH(Txns_i)$ in $MTree$. ii) $Bloom~Proof=\{Bloom~Filter_k|any~Bloom~Filter$ $that~``error~contains"~account~i\}$\footnote{Due to the possibility of false positives in a bloom filter, indicating that elements (account addresses) not belonging to the target set may be included, it is necessary to furnish information that can completely reconstruct this bloom filter (i.e., all elements of the target set) as evidence.}. The ``Block Index" corresponds to the index number of the block associated with the relevant ``Proof". The visual representation of the logical relationship in the VPB pair data structure is depicted in Figure~\ref{fig:VPB pair}.
    \begin{figure*}[htp!]
        \centering
        \includegraphics[width=0.92\linewidth]{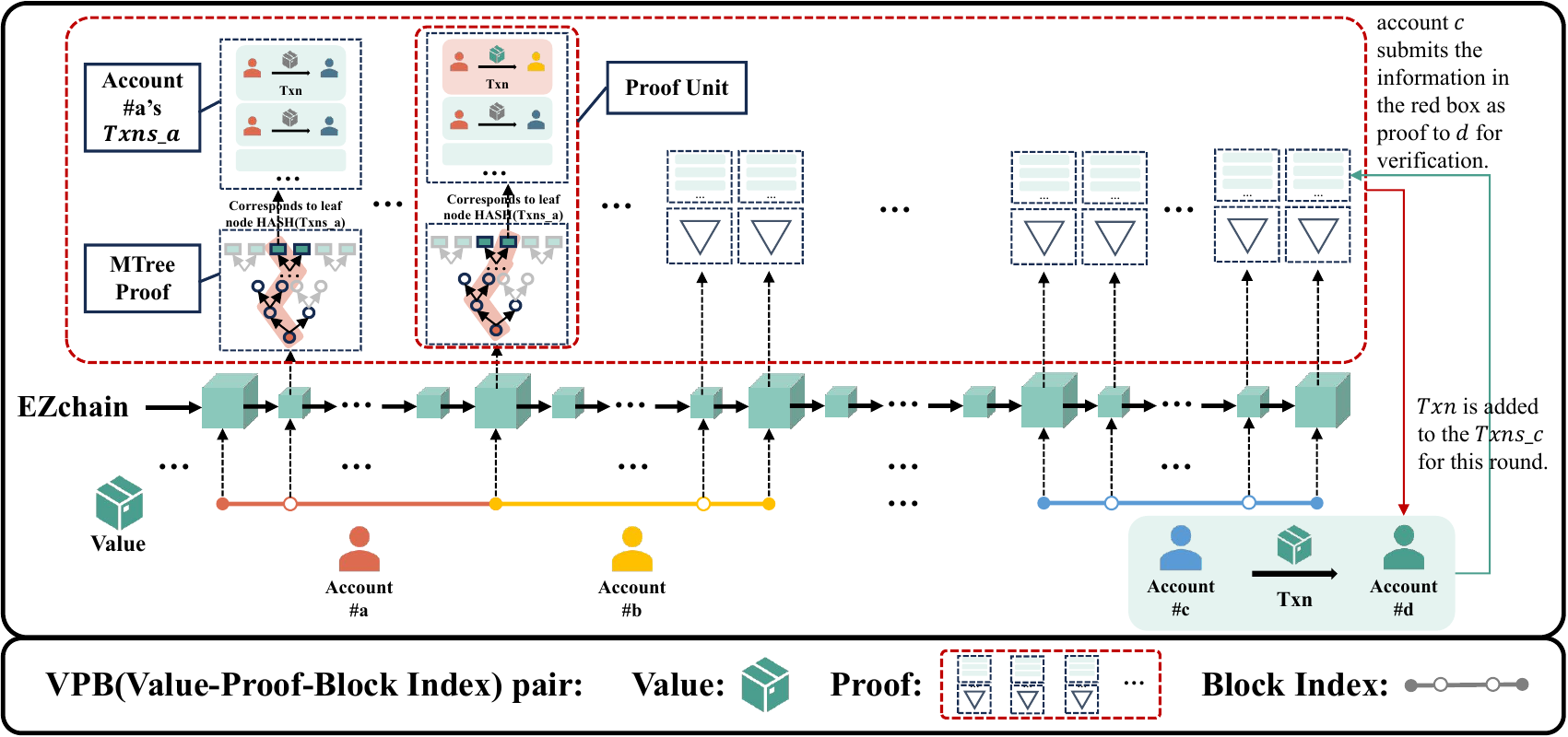}
        \caption{VPB pair's data structure in EZchain.}
        \label{fig:VPB pair}
    \end{figure*}
\end{enumerate}

\subsection{Analysis of specific transaction cases in EZchain}
\label{subsec: EZchain case}

This subsection provides a detailed analysis of the involvement of each node in the EZchain system during various stages of a transaction, including initiation, submission, consensus, and confirmation. We will also explain the operational process of the EZchain system using the example depicted in Figure~\ref{fig:VPB pair}.

Assuming that the value, green cargo box in Figure~\ref{fig:VPB pair}, is allocated to account $a$ in the genesis block (block height $0$). Subsequently, account $a$ intends to submit transaction $Txn_{\alpha}=(a, b, value, Time, SigInfo)$ to a block (block height $\alpha$), for transferring the aforementioned value to account $b$. In this case, account $a$ needs to perform the following steps: i) Add $Txn_{\alpha}$ to the pending transaction set $Txns_a$ and subsequently submit $AccTxn_a=(Sender_a, HASH(Txns_a), SigInfo_a)$ to the transaction pool.
ii) Wait for the miner to broadcast block $\alpha$ and provide the corresponding $MTree~Proof$ of $AccTxn_a$ to account $a$.
iii) Account $a$ transfers the VPB pair associated with the value to $b$.
iv) Account $b$ verifies the VPB pair received from $a$. Upon successful verification, $b$ confirms $Txn_{\alpha}$.

In the above case, the VPB pair associated with the value should contain the following proof: all information of $Txns_a$ within blocks from height $0$ to $\alpha$, along with their corresponding $MTree~Proof$s. This combination is referred to as $Proof~Unit=(Txns_a, MTree~Proof)$. Multiple $Proof~Unit$s are combined into the $Proof^a_b$ that account $a$ submits to account $b$. It's important to note that the last $Proof~Unit$ in $Proof^a_b$ should include $Txn_{\alpha}$. The $Block~Index$ in the VPB pair represents the block index numbers corresponding to all the $Proof~Unit$s in $Proof^a_b$.

In the verification process, we provide an informal but intuitive approach for account $b$ to verify the VPB pair provided by account $a$. Initially, account $b$ needs to ensure the accuracy of the $Block~Index$.This can be achieved by utilizing the EZchain main chain's $Bloom~Filter$. Since the main chain is established through network consensus, the $Bloom~Filter$ remains consistent and accurate among all nodes. Except for some rare certain cases\footnote{Consensus nodes may repeatedly add the addresses and transactions of honest accounts to the Bloom filter and Mtree in the EZchain block. As a result, EZchain allows honest accounts to issue ``challenges" to the consensus node responsible for publishing the corresponding block. These challenges require the consensus node to provide relevant proof, specifically the necessary information required to reconstruct the Bloom filter or Mtree, in order to resolve the challenge. Legitimate proof can demonstrate that honest account nodes either suffered from false positives in the Bloom filter or were redundantly submitted by the consensus node.}, $b$ can utilize the $Bloom~Filter$ to determine the blocks in which $a$ has submitted transactions. Subsequently, based on the correct $Block~Index$, $b$ can employ the corresponding $Proof~Unit$s to verify whether $a$ has utilized the value within any of the transactions submitted across blocks from height $0$ to $\alpha$. If $b$ discovers that the value has been transferred prior to $Txn_{\alpha}$, then $Txn_{\alpha}$ is deemed an illegal transaction, and $b$ should reject it. Otherwise, the transaction can be validated as legitimate.

The aforementioned verification process requires $b$ to combine information from the main chain and the proof provided by $a$ to jointly verify the legitimacy of the transaction. The on-chain information of EZchain ensures that $a$ cannot provide false or tampered ``value's history". Therefore, $a$ can only provide truthful, complete, and verifiable proof to $b$.

If $b$ confirms $Txn_{\alpha}$, it then becomes the new owner of the value. Later, if $b$ wants to transfer the value to account $c$ in block $\beta$ ($\beta > \alpha$), the operation is similar to when $a$ provided the VPB pair to $b$ earlier, except that the $Proof$ and $Block~Index$ now need to cover all historical information of both $a$ and $b$ (i.e., all $a$'s transactions within blocks $0$ to $\alpha$, and all $b$'s transactions within blocks $\alpha$ to $\beta$). Moreover, EZchain fully supports transactions involving multiple values, so the aforementioned operation needs to be performed for each included value.

\subsection{Optimized design of EZchain}
\label{subsec: Optimized design}

\subsubsection{Selection mechanism of values during transaction}
\label{subsubsec: Selection mechanism of values}
\begin{figure}[htp!]
    \centering
    \includegraphics[width=1\linewidth]{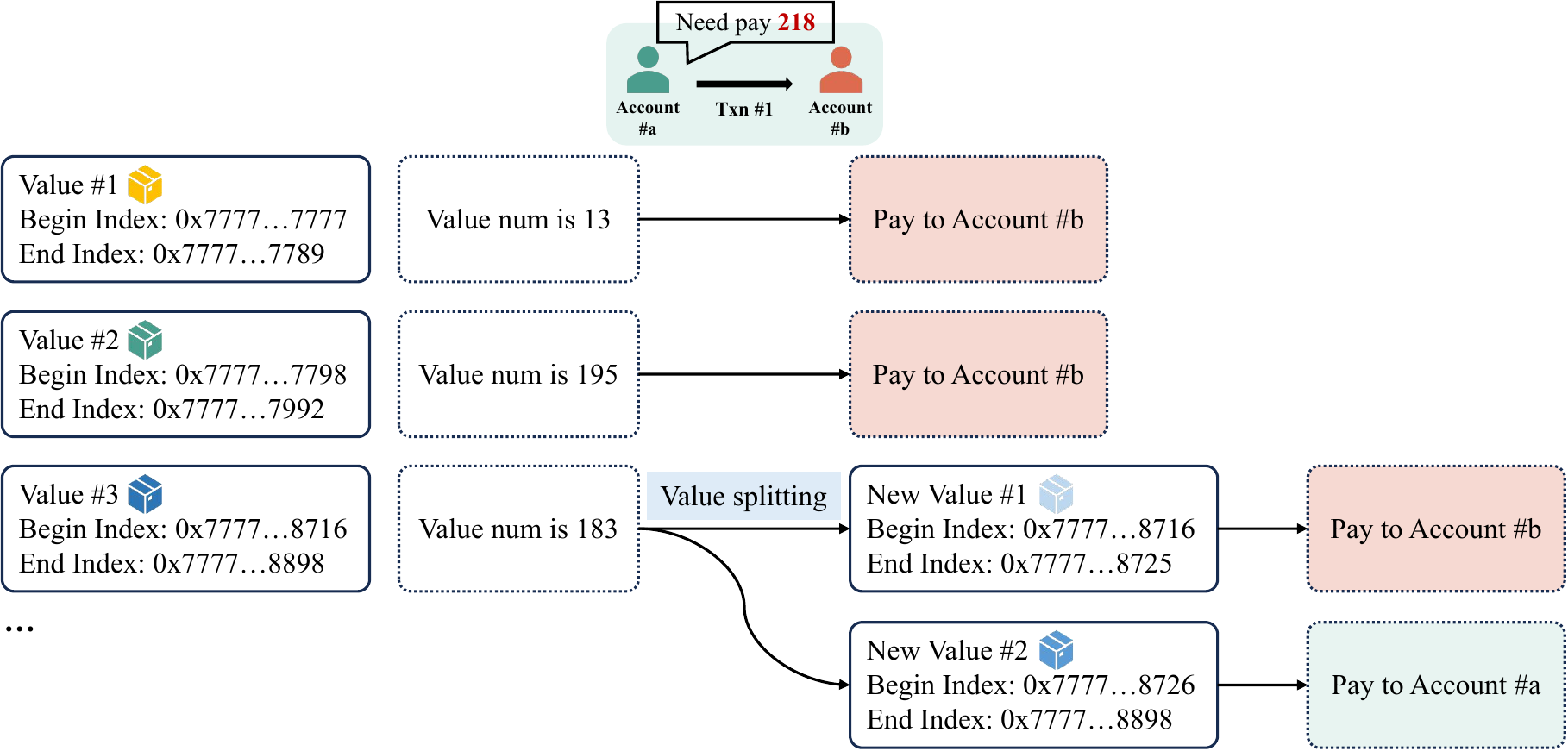}
    \caption{Unoptimized selection mechanism of transaction's value set.}
    \label{fig: unoptimized selection plan}
\end{figure}

During the transaction, assuming that account $a$ wants to pay a target amount $x$ to account $b$, then $a$ needs to collect the target amount from the values it holds. The simplest way is to iterate through the values from the beginning until finding the values whose cumulative amount satisfies $x$ (as shown in Figure~\ref{fig: unoptimized selection plan}).

\begin{figure}[htp!]
    \centering
    \includegraphics[width=1\linewidth]{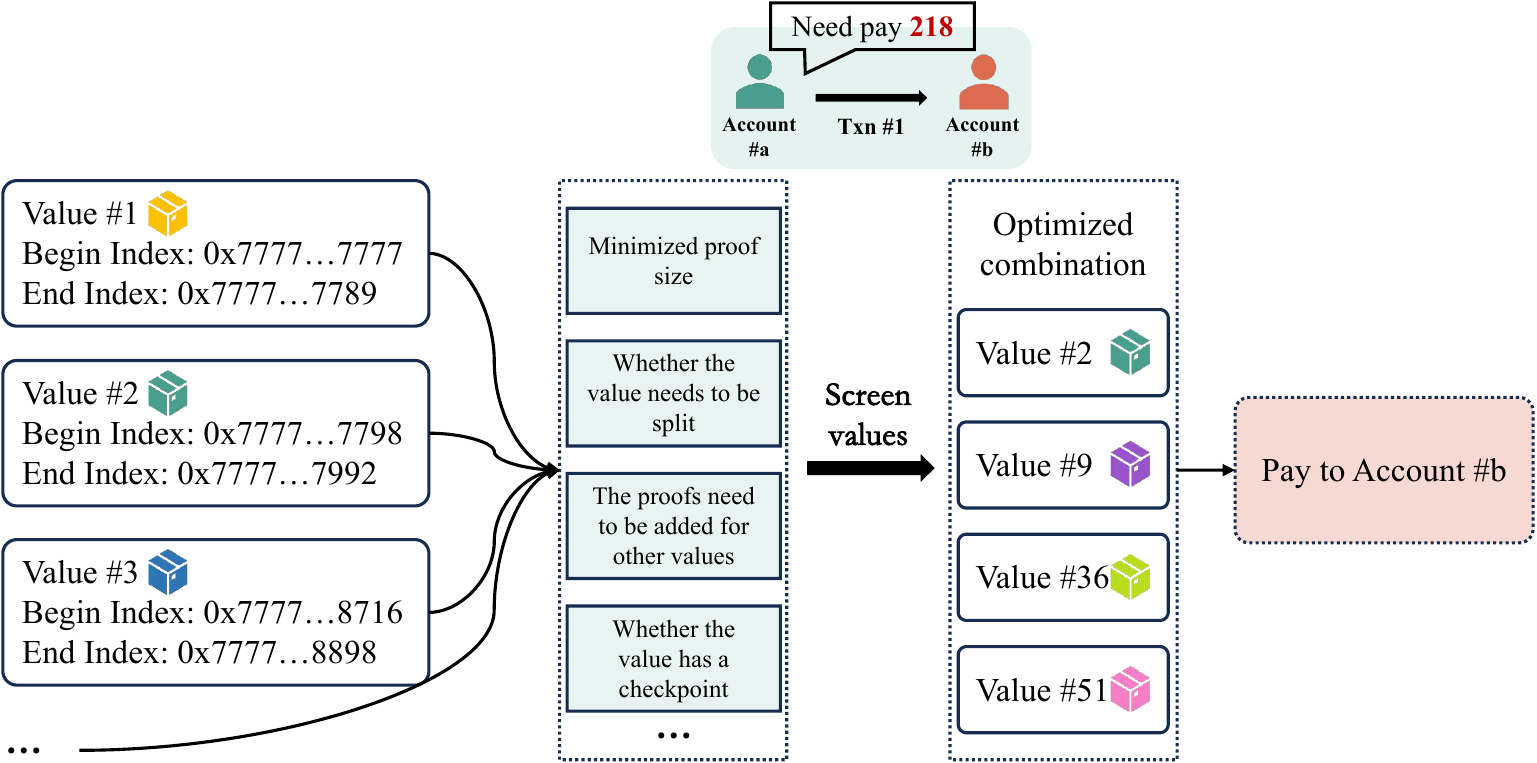}
    \caption{Optimized selection mechanism of transaction's value set.}
    \label{fig: optimized selection plan}
\end{figure}

However, the account node can analyze the specific situation and select the value set that is most suitable for this transaction. For example, $a$ can choose the value set that requires the smallest proof, or does not require splitting (change), etc. (as shown in Figure~\ref{fig: optimized selection plan}), after comprehensive consideration, each account can customize its own optimization selection strategy according to its own situation.

\subsubsection{Check point mechanism}
\label{subsubsec: Check point mechanism}
\begin{figure*}[h!]
    \centering
    \includegraphics[width=0.92\linewidth]{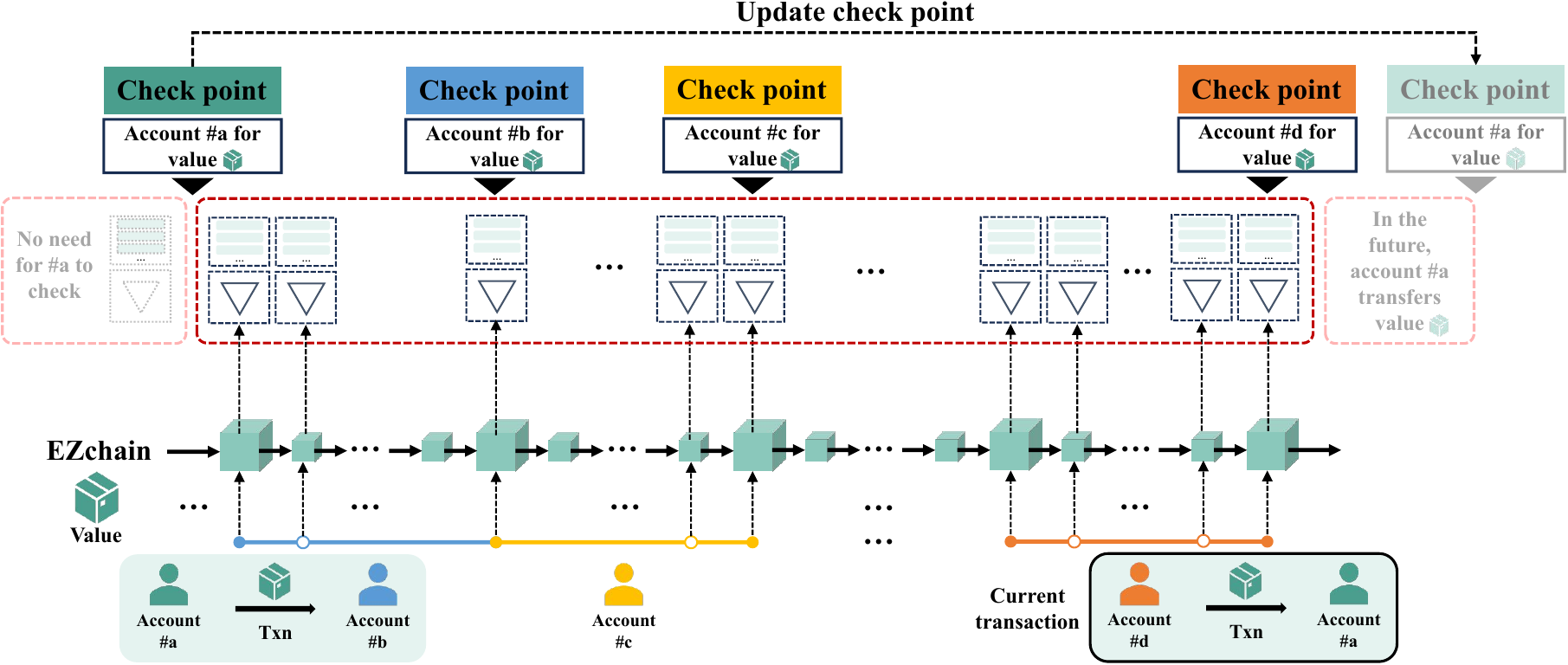}
    \caption{EZchain's check point mechanism.}
    \label{fig: EZchain check point mechanism}
\end{figure*}
The requirement to indiscriminately request all transfer records and proofs for any given value from an account is not necessary. Here is a simple and intuitive example to illustrate (as shown in Figure~\ref{fig: EZchain check point mechanism}): In block $\alpha$, $a$ transfers a certain value number (e.g., $100$) to $b$. In a subsequent block $\beta$ ($\beta > \alpha$), $d$ wants to transfer a value of $100$ back to $a$. In this case, $d$ can choose the value that was previously transferred from $a$ and inform $a$ that this value does not need to be revalidated before block $\alpha$, as it has already been locally validated by $a$. In other words, $a$ only needs to verify the legality of the value within the block height ranging from $\alpha + 1$ to $\beta$. This design concept constitutes the foundation of EZchain's check point mechanism. As the transactions progress, accounts can continuously update the check points for the held values to save on storage, communication, and verification costs.

\section{Analysis of EZchain's performance, security and decentralization}
\label{sec: EZchain's analysis}

\subsection{Analysis of EZchain's performance}
\label{subsec: EZchain's performance analysis}
\begin{figure}[htp!]
    \centering
    \includegraphics[width=1\linewidth]{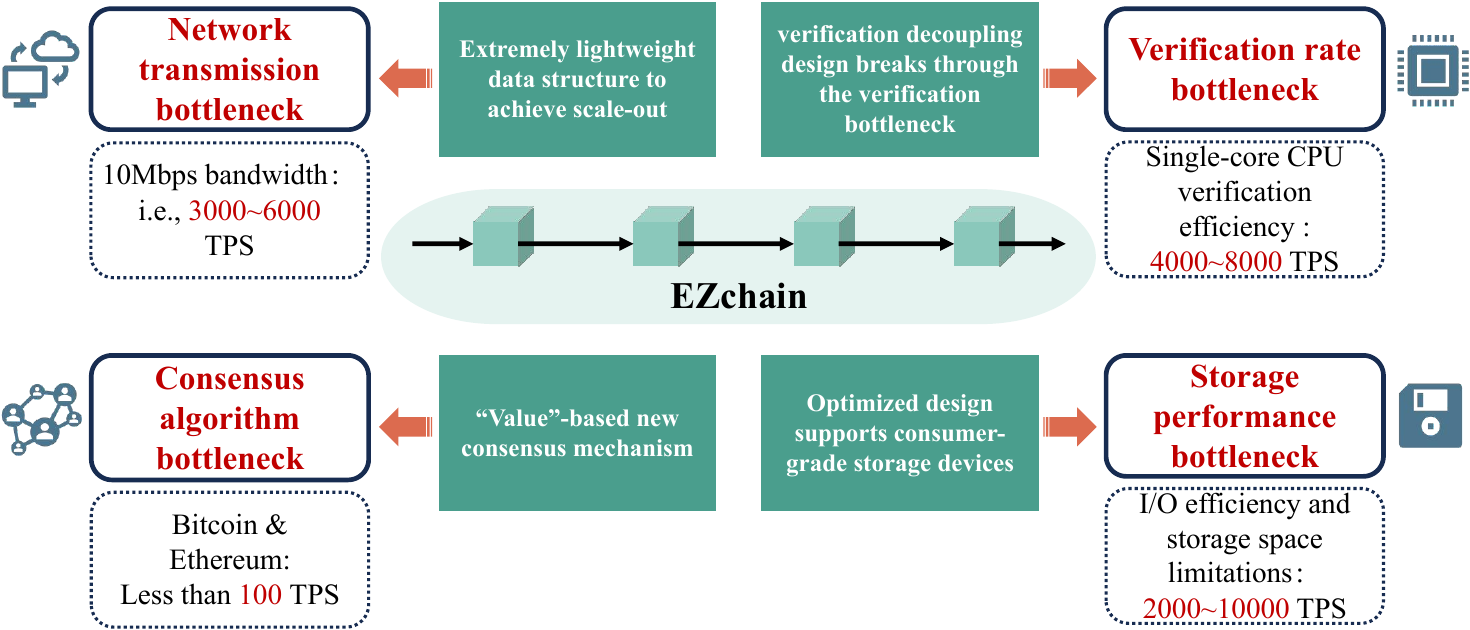}
    \caption{EZchain's performance breakthroughs in consensus, network, verification and storage.}
    \label{fig: EZchain's performance breakthrough}
\end{figure}
As depicted in Figure~\ref{fig: EZchain's performance breakthrough}, EZchain has made significant progress in multiple performance bottleneck indicators.

In terms of consensus and communication, the current majority of layer-1 blockchain still struggle to achieve ideal practical efficiency. In general, the most outstanding works can achieve $100\%$ bandwidth utilization in theory or nearly $10,000$ TPS in expensive bandwidth environments. However, strictly speaking, they have not achieved ``scale-out" because when the system reaches the physical limit of bandwidth, adding more nodes and transaction submissions will only worsen the network congestion. Although layer-2 protocols, permissioned chains, and private chains can address the aforementioned dilemma, the former's security is almost decoupled from the main chain, and the high entry barrier of the latter is unfavorable for decentralization.

\begin{figure}[h!]
    \centering
    \includegraphics[width=1\linewidth]{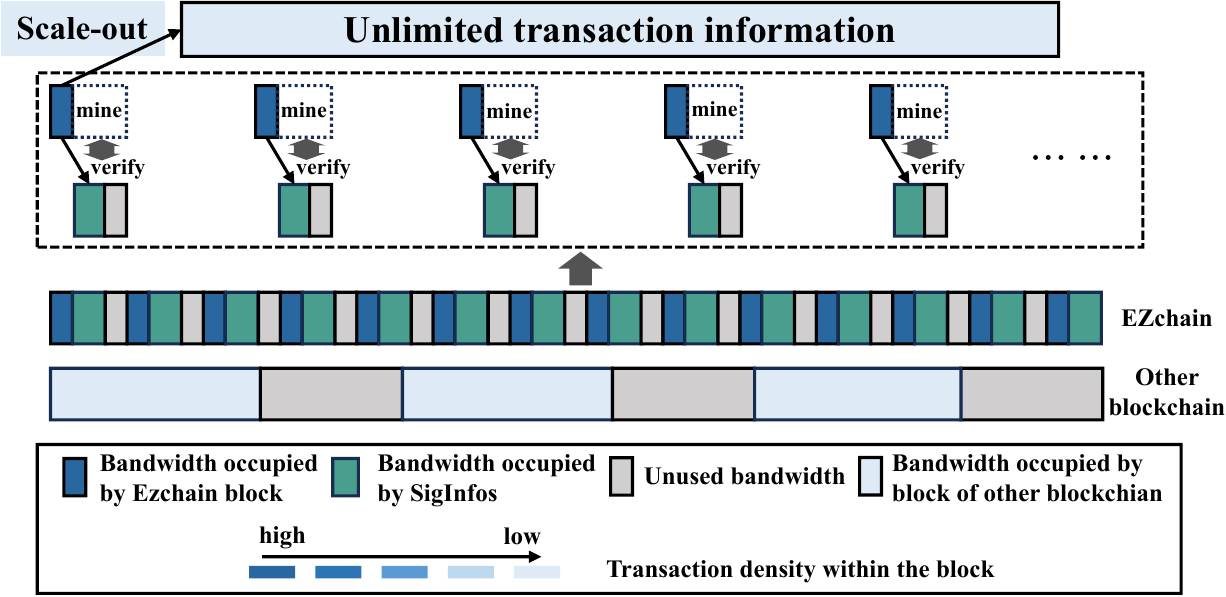}
    \caption{Comparison of consensus and communication efficiency between EZchain and other blockchain.}
    \label{fig:communication cost}
\end{figure}

EZchain ensures the atomicity and legality of all transactions by only relying on the consensus of the Merkle tree root and the Bloom filter information. As shown in Figure~\ref{fig:communication cost}, this enables the following benefits: i) The size of the Merkle tree root and Bloom filter will not change with network size and accumulation of transactions, therefore, EZchain block have a constant size; ii) And theoretically accommodating an unlimited number of transactions within the fixed-sized block. These two points collectively contribute to achieving ``scale-out" at the consensus and communication levels in EZchain.

In terms of storage cost, due to the special design of EZchain, the block information agreed upon by consensus nodes only includes the Merkle tree root, Bloom filter, random number nonce, timestamp, miner's signature, miner's address, and block's index. Additionally, each EZchain block has a nearly constant size (approximately 0.5 Mb) that does not change with the increase in system throughput and nodes' scale.\footnote{Furthermore, consensus nodes also need to store complete Merkle tree information corresponding to recent blocks to respond to ``challenges" from other nodes.} This design significantly reduces the storage cost for consensus nodes. For account nodes, they only need to store relevant information about their own holdings (i.e., values, VPB pairs, checkpoints). Moreover, based on subsequent simulation experiments (Subsection~\ref{subsec: Storage and verification cost acc}), it can be observed that these storage costs increase very slowly with the system's runtime. What's even more exciting is that using reasonable optimization schemes (for example, the selection mechanism of values mentioned in Subsection~\ref{subsec: Optimized design}), the amount of proof information required for peer-to-peer transmission converges to a fixed value as the system runs (experiment in Figure~\ref{fig: ex Storage and verification cost acc 3}). This means that the transmission and storage costs for account nodes are essentially of constant magnitude, without increasing with the number of nodes and transactions. In conclusion, at the storage level, all nodes in the EZchain system achieve ``scale-out".

In terms of verification costs, the validation cost of consensus nodes is almost constant and does not increase significantly with system throughput and network size. The validation cost of account nodes is also almost constant due to the checkpoint mechanism. Both of these points can be well verified in subsequent simulation experiments.

\begin{figure*}[h!]
    \centering
    \includegraphics[width=0.95\linewidth]{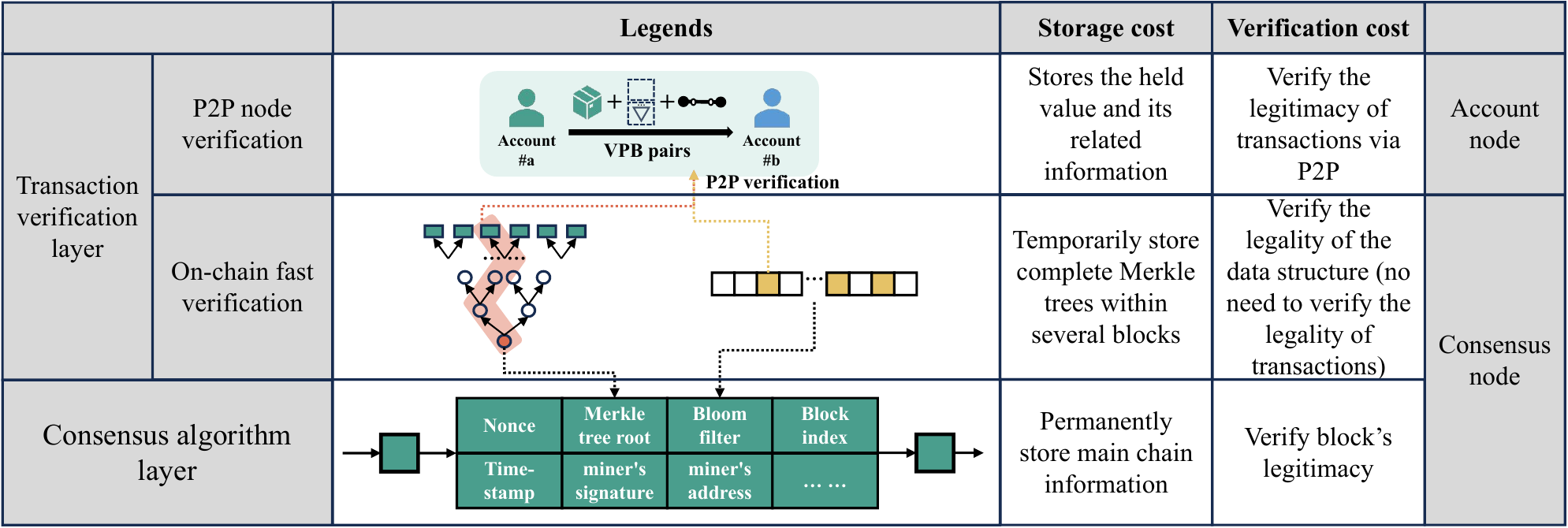}
    \caption{Storage and verification costs of each node in EZchain.}
    \label{fig: Storage and verification costs}
\end{figure*}

The visualization of all nodes' storage and verification costs are shown in Figure~\ref{fig: Storage and verification costs}.

\subsection{Analysis of EZchain's security and decentralization}
\label{subsec: EZchain's security and decentralization analysis}

Regarding security, although EZchain has implemented significant alterations to the data structure compared to traditional blockchains, the backbone consensus mechanism of its main chain remains unchanged. During the P2P verification phase for account nodes, EZchain's algorithm guarantees the result certainty.  Additionally, all proof transfers occur in a point-to-point manner (between two account nodes or from consensus node to account node), maintaining the network assumptions of the customized backbone consensus algorithm (e.g., asynchrony, weak synchrony and synchrony).

Concerning decentralization, EZchain has effectively lowered various ``thresholds", enabling broader node participation in the system and promoting decentralization. For example, the verification calculation complexity of the consensus node persists at a constant level. Furthermore, the storage and message complexity can approach equivalence to Bitcoin.\footnote{EZchain accomplishes this by adjusting the block interval to 10 minutes, reasonably sacrificing a degree of transaction confirmation latency to substantially reduce storage costs without influencing system throughput.}

Consequently, EZchain retains robust security guarantees stemming from the customized backbone consensus algorithm while enhancing decentralization to some degree.

\section{Simulation experiments of EZchain}
\label{sec: Simulation experiments}

In this section, we assess performance by implementing a simulation prototype of EZchain (including the network transport layer) and concentrating primarily on the subsequent aspects: i) The average throughput of the EZchain system; ii) Storage consumption and verification time for EZchain consensus nodes; iii) Storage consumption and transaction confirmation latency for EZchain account nodes.

\subsection{Experimental setup of EZchain}
\label{subsec: Experimental setup}

\subsubsection{Experimental equipment}
\label{subsubsec: Experimental equipment}


In this experiment, we utilized a hardware platform equipped with an 11th Gen Intel(R) Core(TM) i7-1165G7 processor (2.80GHz, octa-core) and 16.0 GB of memory to conduct simulations for EZchain. The experimental setup varied in node count, with the largest configuration comprising 100 consensus nodes and 180 account nodes. The maximum number of simulation rounds is set to 1200, allowing for a comprehensive assessment of EZchain's performance under large-scale node deployment and extended operational durations. The simulator, written in Python, can be accessed on Github\footnote{https://github.com/Re20Cboy/Ezchain-py}.


\subsubsection{Simulation system setup}
\label{subsubsec: Simulation system setup}

The quantity of consensus nodes and account nodes can be configured flexibly. Given device memory limitations, parameters for this experiment range from $3$ to $160$ nodes. All nodes of EZchain randomly establish P2P connections, with each node having a maximum of 30 neighboring nodes, and the bandwidth is uniformly set to 1 Mbits/s.  Communication latency between consensus nodes (referring here to queuing delays, excluding transmission times, and similarly below) falls within 1 s as a random variable. Communication latency between consensus nodes and account nodes is fixed at 1.5 s, while latency between account nodes is likewise 1.5 s. The genesis block records the ownership of all initial values. In each simulation round (block generation), account nodes spontaneously engage in random transactions, with the transaction amounts adjustable based on the experiment.

\subsection{EZchain system throughput test}
\label{subsec: throughput test}

\begin{figure}[htp!]
    \centering
    \includegraphics[width=0.87\linewidth]{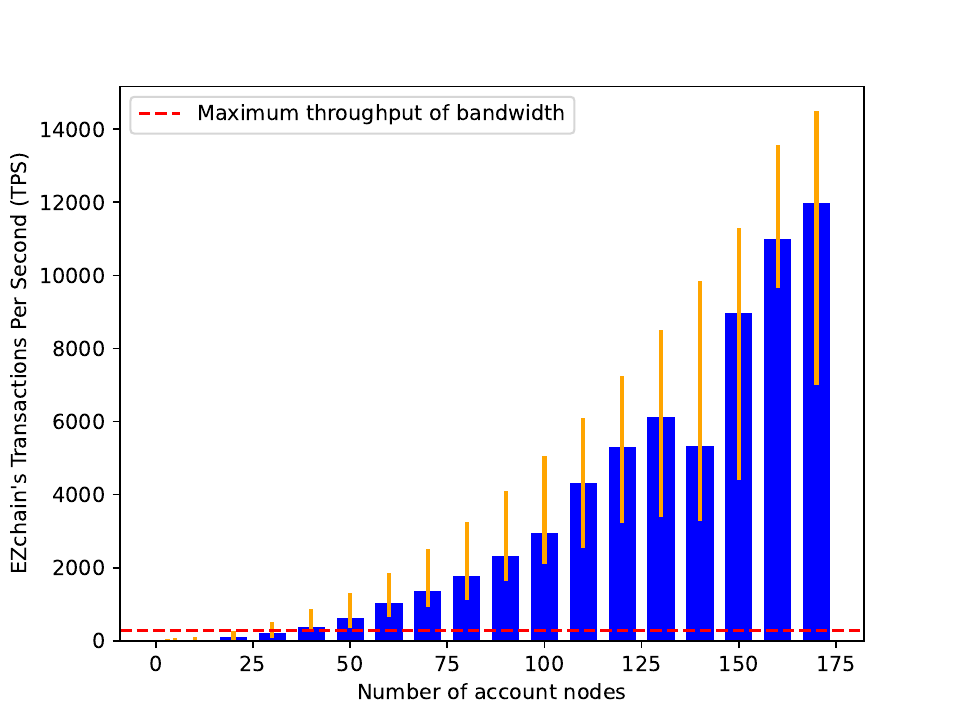}
    \caption{System throughput test of EZchain.}
    \label{fig: ex throughput test 1}
\end{figure}

Due to equipment constraints and to maximize the test of EZchain's throughput limits, we reduce execution rounds and increase extension nodes while enlarging the transaction scale. Additionally, given the quadratic relationship between account nodes and injected transactions per round\footnote{Here we assume at most one unidirectional transaction can be submitted between two accounts per round (block). Therefore, a maximum of $n^2$ transactions can occur with $n$ account nodes.}, we configure 100 consensus nodes and $3$ to $160$ account nodes, with throughput test results shown in Figure~\ref{fig: ex throughput test 1}. The red dotted line is the throughput of the classic blockchain with $100\%$ utilization of bandwidth resources, and the yellow bar indicates the floating range of the test results. Results demonstrate EZchain can achieve over 10,000 Transactions Per Second (TPS), with each block readily accommodating approximately 15,000 transactions. The system throughput can exceed the bandwidth's limit to meet ``scale-out".

Ultra-high throughput does not affect other performances. At over 10,000 TPS, storage requirements for each blocks less than 0.5 Mb, which are lower than the block size in Bitcoin, despite thousands-fold higher throughput versus Bitcoin. For validation time, EZchain's consensus node block verification occurs primarily within the order of 1 millisecond, while the account node's transaction confirmation delay also approaches 10 seconds.

\subsection{Storage and verification cost of EZchain consensus node}
\label{subsec: Storage and verification cost}

\begin{figure}[htp!]
    \centering
    \includegraphics[width=0.87\linewidth]{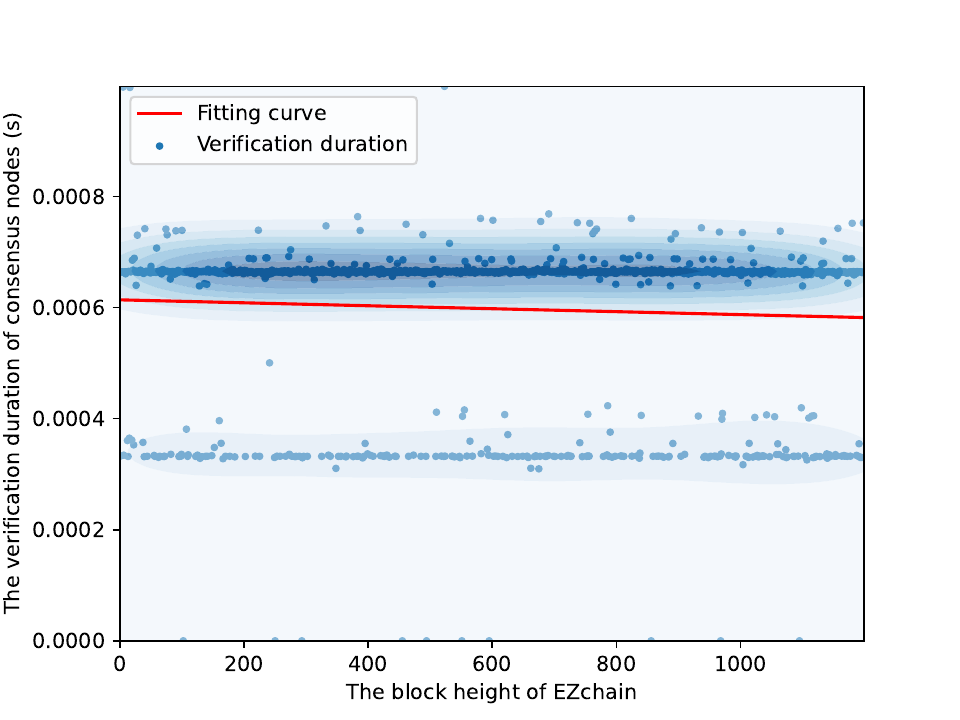}
    \caption{Verification cost of EZchain consensus node.}
    \label{fig: ex Validation cost 1}
\end{figure}

Given device constraints, and to extensively test consensus node storage and validation overheads, we require expanded experimental rounds (system runtime) to observe long-term costs. Consequently, this experiment utilizes 3 consensus nodes and 3 account nodes across 1,200 rounds (blocks) to assess cost trends for EZchain consensus nodes. Results in Figure~\ref{fig:  ex Storage cost 1} that storage requirements approximate 150 Mb across 1,200 blocks with a linear growth trend. As shown in Figure~\ref{fig: ex Validation cost 1}, validation times per block remain entirely independent of system throughput, transaction count, and runtime, persisting below 1 millisecond.

\begin{figure}[htp!]
    \centering
    \includegraphics[width=0.87\linewidth]{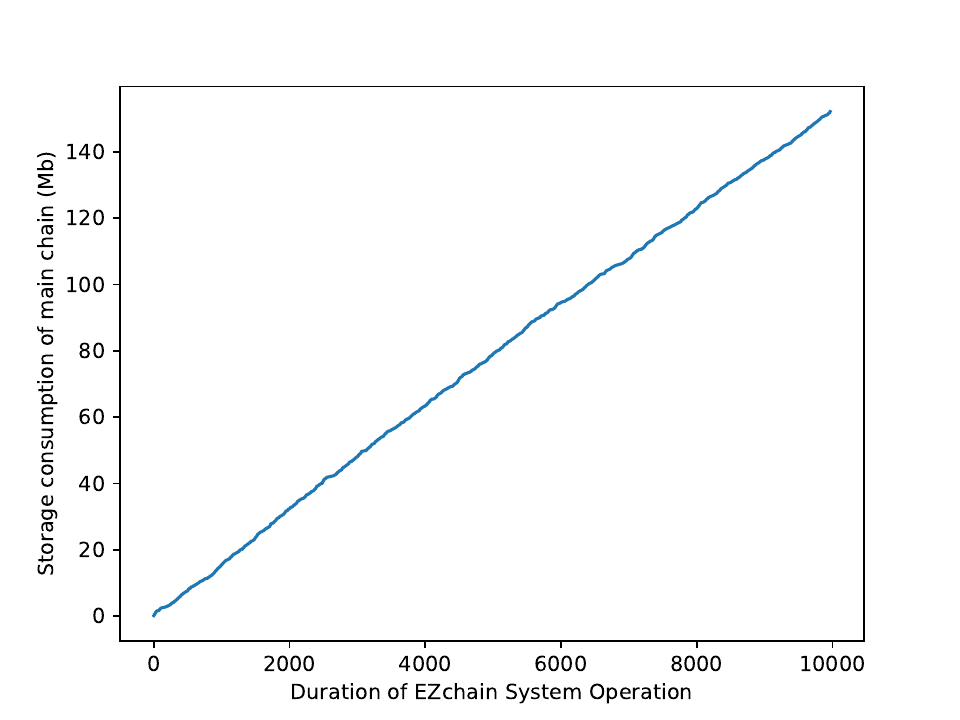}
    \caption{Storage cost of EZchain main chain.}
    \label{fig: ex Storage cost 1}
\end{figure}

\subsection{Storage and verification cost of EZchain account node}
\label{subsec: Storage and verification cost acc}

\begin{figure}[htp!]
    \centering
    \includegraphics[width=0.87\linewidth]{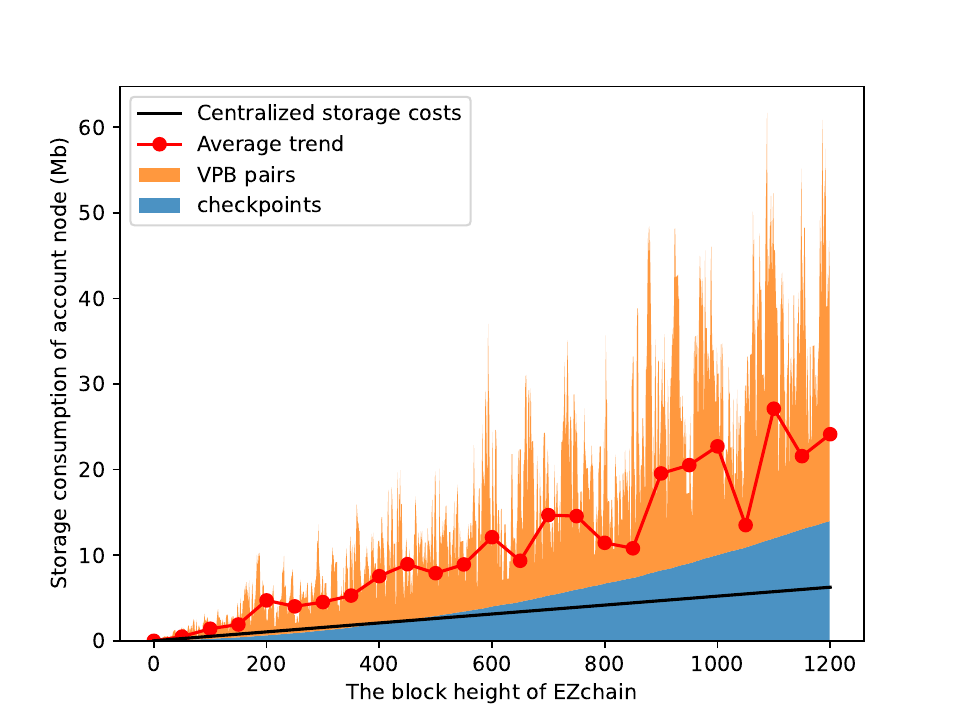}
    \caption{Storage cost of account nodes.}
    \label{fig: ex Storage and verification cost acc 1}
\end{figure}

In this experiment, we establish 3 consensus nodes and 3 account nodes, executing 1200 rounds (blocks) to monitor the consumption trends of EZchain account node's storage cost. The results are depicted in Figure~\ref{fig: ex Storage and verification cost acc 1}, where the red line represents the smoothed cost (an average value is taken every 50 rounds), and the black line represents the amount of information that the account node needs to store under the same centralized architecture. The average trends indicates that the storage cost of account nodes does not exhibit a substantial increase as the number of system rounds rises. Furthermore, the storage cost does not exhibit an order of magnitude difference compared to that of centralized account nodes.

\begin{figure}[htp!]
    \centering
    \includegraphics[width=1.12\linewidth]{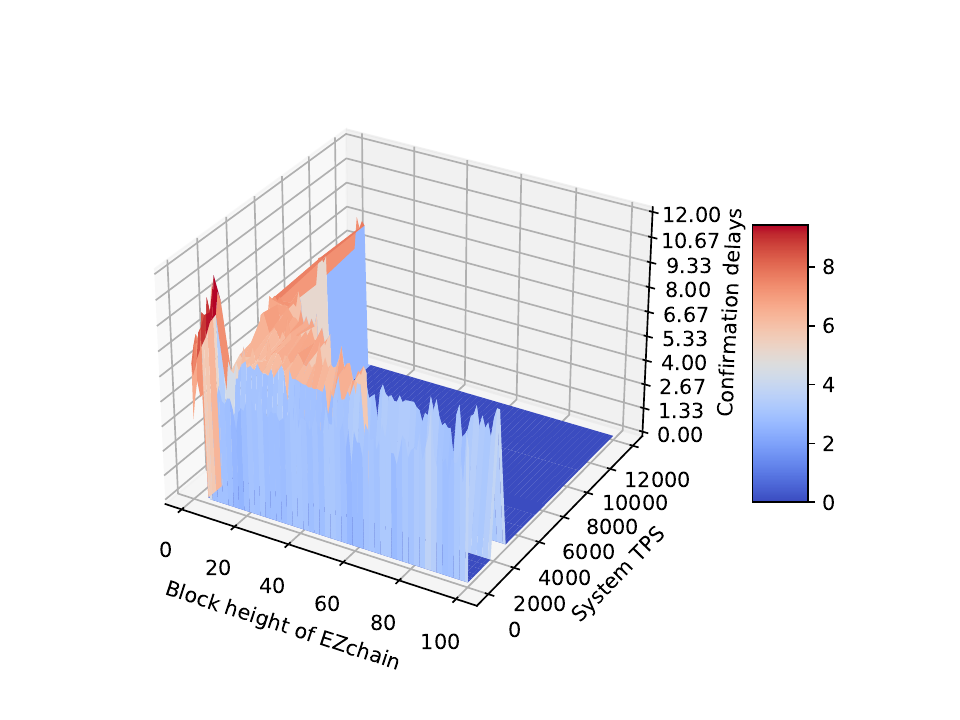}
    \caption{Transaction confirmation delay of account node (deep blue color indicates data absence).}
    \label{fig: ex Storage and verification cost acc 2}
\end{figure}

Regarding transaction confirmation delay, the results, as shown in Figure~\ref{fig: ex Storage and verification cost acc 2}, demonstrate that the transaction confirmation delay for account nodes predominantly falls within 10 seconds. Besides, this delay does not change with the system throughput or system running rounds.

\begin{figure}[htp!]
    \centering
    \includegraphics[width=0.87\linewidth]{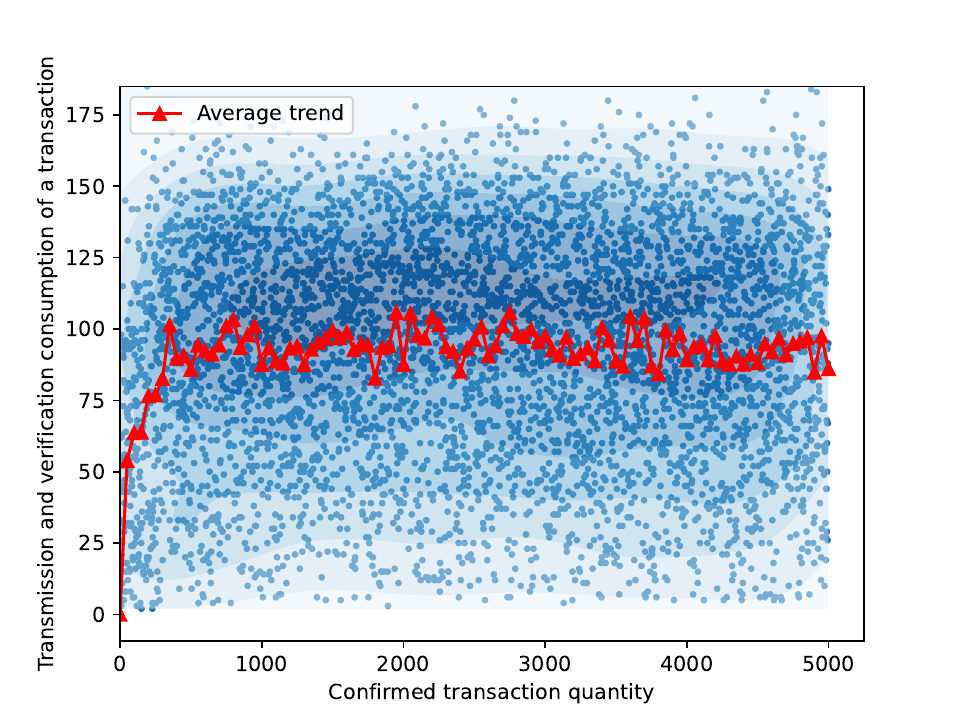}
    \caption{Trend analysis of cumulative transmission and verification cost of transaction.}
    \label{fig: ex Storage and verification cost acc 3}
\end{figure}

For the transmission and verification costs of account nodes, we assume that account nodes randomly conduct transactions at a frequency with a fixed expectation. This allows us to conveniently configure specific experiments, and reflect the transmission and validation costs by tracking the number of holders the value goes through on each transfer. The results, as shown in Figure~\ref{fig: ex Storage and verification cost acc 3}, demonstrate that as transactions progress, the amount of information required for transmission and validation by the account nodes converges to a fixed value. Intuitively, this conclusion shows that the length of the VPB (the width of the red dotted box) in Figure~\ref{fig:VPB pair} will converge to a fixed value, which strongly supports EZchain's scale-out in terms of account node.

\begin{table*}[h]
\caption{Performance Comparison between EZchain and other solutions.}
\centering
\begin{tabular}{@{}cc|ccc|ccc@{}}
\toprule
\multicolumn{2}{c|}{\multirow{2}{*}{}} & \multicolumn{3}{c|}{Experimental parameters}                                                                                                                                                            & \multicolumn{3}{c}{System performance}                                                                                                 \\ \cmidrule(l){3-8} 
\multicolumn{2}{c|}{}                  & Resiliency                                                                                                        & Network size & \begin{tabular}[c]{@{}c@{}}Available\\ bandwidth (Mbps)\end{tabular} & Throughput                  & Latency                 & \begin{tabular}[c]{@{}c@{}}Storage cost per node\\ after 30k txns\end{tabular} \\ \midrule
\multicolumn{2}{c|}{Algorand~\cite{Algorand}}          & t \textless n/3                                                                                                   & 50,000 nodes  & 10 Mbps                                                              & 900 tps                     & 60 sec                  & 16.1MB                                                                         \\
\multicolumn{2}{c|}{Elastico~\cite{LuuNZBGS16}}          & t \textless n/4                                                                                                   & 1,600 nodes  & 20 Mbps                                                              & 40 tps                      & 800 sec                 & 14.4 MB                                                                        \\
\multicolumn{2}{c|}{OmniLedger~\cite{OmniLedger}}        & t \textless n/4                                                                                                   & 1,800 nodes  & 20 Mbps                                                              & 3,500 tps                   & 63 sec                  & 4.5 MB                                                                         \\
\multicolumn{2}{c|}{RapidChain~\cite{RapidChain}}        & t \textless n/3                                                                                                   & 4,000 nodes  & 20 Mbps                                                              & 7,380 tps                   & 8.7 sec                 & 0.92 MB                                                                        \\
\multicolumn{2}{c|}{OHIE~\cite{OHIE}}              & t \textless n/2                                                                                                   & 12,000 nodes & \begin{tabular}[c]{@{}c@{}}8 Mbps to\\ 20 Mbps\end{tabular}          & 2,400 tps                   & 200 sec                 & -                                                                              \\
\multicolumn{2}{c|}{S-HS~\cite{GaiNB0W23}}              & t \textless n/3                                                                                                   & 128 nodes    & 100 Mbps                                                              & 20,000 tps                   & 3 sec                   & -                                                                              \\
\multicolumn{2}{c|}{Conflux~\cite{conflux}}           & t \textless n/2                                                                                                   & 10,000 nodes  & 40Mbps                                                               & 3,200 tps                    & 600 sec                 & 14.3 MB                                                                        \\
\multirow{2}{*}{EZchain}   & con-node  & \multirow{2}{*}{\begin{tabular}[c]{@{}c@{}}t \textless n/2 (PoW type) or\\ t \textless n/3 (BFT type)\end{tabular}} & 100 nodes     & \multirow{2}{*}{1Mbps}                                               & \multirow{2}{*}{10,300+ tps} & \multirow{2}{*}{10 sec} & 0.06 MB                                                                        \\
                           & acc-node  &                                                                                                                   & 180 nodes     &                                                                      &                             &                         & 22.75 MB                                                                       \\ \bottomrule
\end{tabular}
\label{table: Performance Comparison}
\end{table*}

Furthermore, based on the results of this experiment, we can formally prove an important conclusion: 

\begin{theorem}
For an EZchain account node $Acc$:
\begin{enumerate}
\item the quantity of values $Acc$ holds (denoted as $N_v$),
\item $Acc$'s transaction frequency (denoted as $F_{txn}$), and
\item $Acc$'s average duration of holding a value (denoted as $D_v$),
\end{enumerate}
when the above three conditions all have constant upper bounds, the storage cost of $Acc$ will have a fixed constant upper bound, which remains unaltered with respect to the system's running time, cumulative transactions, network size, and other factors.
\label{Conclusion: acc storage cost}
\end{theorem}

\begin{proof}
Based on the design of EZchain, the storage cost of $Acc$ mainly comprises of: i) the VPBs associated with the current values held by $Acc$ and ii) the checkpoints ($CKs$), while disregarding the negligible probability of bloom proof occurrence. In formal terms, $Acc_{storage~cost} = VPBs_{storage~cost} + CKs_{storage~cost}$, where $VPBs_{storage~cost} = N_v*CK_{gap}*D_v*F_{txn}*S_{pu}$. $CK_{gap}$ represents the number of holders containing this value in VPB, i.e., the ordinate of Figure~\ref{fig: ex Storage and verification cost acc 3}. Intuitively, $CK_{gap}$ is equivalent to the number of holders between two green checkpoints in Figure~\ref{fig: EZchain check point mechanism}. Figure~\ref{fig: ex Storage and verification cost acc 3} demonstrates the convergence of $CK_{gap}$. $S_{pu}$ denote the size of a single proof unit, which is also a constant value. Hence, $VPBs_{storage~cost}$ represents a fixed cost.

As shown in Figure~\ref{fig: ex Storage and verification cost acc 1}, the growth of $CKs_{storage~cost}$ is extremely slow, remaining largely on par with the consumption of account nodes in centralized transaction systems. However, unlike the latter, due to the fixed quantity of values in EZchain, $CKs_{storage~cost}$ also has a theoretical upper bound. On the other hand, centralized transaction storage continuously increases with the accumulation of transactions.
\end{proof}

\subsection{Performance Comparison between EZchain and other solutions}
\label{subsec: Performance Comparison}

The comparative analysis concentrates on EZchain in comparison to other layer-1 solutions. Although most layer-2 solutions demonstrate remarkable performance, they function independently from the underlying blockchain and necessitate more robust security assumptions for transactions. The comparative results, illustrated in Table~\ref{table: Performance Comparison}, emphasize our main considerations: system throughput, transaction confirmation latency, and node storage costs. The data pertaining to other solutions primarily originates from respective experimental sources, with certain data subjected to reasonable scaling adjustments to facilitate improved comparisons.

EZchain demonstrates the second-highest throughput while utilizing minimal network size and the lowest bandwidth resources. Notably, S-HS necessitates 100 times more bandwidth resources compared to EZchain, yet achieves only double the throughput. Moreover, Subsection~\ref{subsec: throughput test} validates EZchain's ability to meet scale-out within limited bandwidth constraints. There are compelling indications that with enhanced experimental infrastructure (such as significantly increased memory resources), EZchain could potentially outperform S-HS entirely. Notably, EZchain consistently delivers confirmation latency within practical levels suitable for consumer-grade applications.

When considering storage costs, EZchain's consensus nodes exhibit a substantially lower storage consumption, at least an order of magnitude less when compared to the optimal sharding solution, as detailed in the provided table. While the storage cost for account nodes is relatively high, Conclusion~\ref{Conclusion: acc storage cost} provides evidence that this cost converges to a fixed value, unlike in other solutions where it grows indefinitely with increasing transaction volumes.

\section{Conclusion}
\label{sec: Conclusion}

In this paper, we introduced EZchain, a novel decentralized distributed ledger blockchain system tailored for Web3.0 applications. A prototype simulation system has been developed and is available as open-source\footnote{https://github.com/Re20Cboy/Ezchain-py}. The empirical evidence suggests that EZchain is capable of achieving ``scale-out" performance, utilizing consumer-grade bandwidth, computational, and storage resources, while maintaining the integrity of decentralization and security principles. Future work will focus on two primary objectives: i) Enhancing EZchain's efficiency by optimizing storage and transmission overhead and integrating advanced algorithmic plugins; ii) Expanding EZchain's design architecture to encompass Turing-complete blockchain systems; iii) Reasonable incentive strategies to build a good ecosystem.

\section*{Acknowledgment}
This work was supported by the National Natural Science Foundation of China (No. 62172385), and the Innovation Program for Quantum Science and Technology (No. 2021ZD0302900).

\newpage
\bibliography{ref.bib}

\begin{thebibliography}{10}
\providecommand{\url}[1]{#1}
\csname url@samestyle\endcsname
\providecommand{\newblock}{\relax}
\providecommand{\bibinfo}[2]{#2}
\providecommand{\BIBentrySTDinterwordspacing}{\spaceskip=0pt\relax}
\providecommand{\BIBentryALTinterwordstretchfactor}{4}
\providecommand{\BIBentryALTinterwordspacing}{\spaceskip=\fontdimen2\font plus
\BIBentryALTinterwordstretchfactor\fontdimen3\font minus
  \fontdimen4\font\relax}
\providecommand{\BIBforeignlanguage}[2]{{%
\expandafter\ifx\csname l@#1\endcsname\relax
\typeout{** WARNING: IEEEtran.bst: No hyphenation pattern has been}%
\typeout{** loaded for the language `#1'. Using the pattern for}%
\typeout{** the default language instead.}%
\else
\language=\csname l@#1\endcsname
\fi
#2}}
\providecommand{\BIBdecl}{\relax}
\BIBdecl

\bibitem{BitcoinWiki2019}
BitcoinWiki, ``Confirmation,'' \url{https://en.bitcoin.it/wiki/Confirmation},
  2019, april, 2019.

\bibitem{VAPOR2018}
\BIBentryALTinterwordspacing
Z.~Ren and Z.~Erkin, ``{VAPOR:} a value-centric blockchain that is scale-out,
  decentralized, and flexible by design,'' \emph{CoRR}, vol. abs/1810.12596,
  2018. [Online]. Available: \url{http://arxiv.org/abs/1810.12596}
\BIBentrySTDinterwordspacing

\bibitem{SpontaneousSharding}
\BIBentryALTinterwordspacing
Z.~Ren, K.~Cong, T.~Aerts, B.~de~Jonge, A.~Morais, and Z.~Erkin, ``A scale-out
  blockchain for value transfer with spontaneous sharding,'' in \emph{Crypto
  Valley Conference on Blockchain Technology, {CVCBT} 2018, Zug, Switzerland,
  June 20-22, 2018}, 2018, pp. 1--10. [Online]. Available:
  \url{https://doi.org/10.1109/CVCBT.2018.00006}
\BIBentrySTDinterwordspacing

\bibitem{Eyal2016BitcoinNG}
I.~Eyal, A.~E. Gencer, and R.~V. Renesse, ``Bitcoin-ng: a scalable blockchain
  protocol,'' in \emph{Usenix Conference on Networked Systems Design and
  Implementation}, 2016.

\bibitem{Kokoriskogias2016Byzcoin}
E.~Kokoriskogias, P.~Jovanovic, N.~Gailly, I.~Khoffi, L.~Gasser, and B.~Ford,
  ``Enhancing bitcoin security and performance with strong consistency via
  collective signing,'' \emph{Applied Mathematical Modelling}, vol.~37, no.~8,
  pp. 5723--5742, 2016.

\bibitem{Algorand}
\BIBentryALTinterwordspacing
Y.~Gilad, R.~Hemo, S.~Micali, G.~Vlachos, and N.~Zeldovich, ``Algorand: Scaling
  byzantine agreements for cryptocurrencies,'' in \emph{Proceedings of the 26th
  Symposium on Operating Systems Principles, Shanghai, China, October 28-31,
  2017}, 2017, pp. 51--68. [Online]. Available:
  \url{https://doi.org/10.1145/3132747.3132757}
\BIBentrySTDinterwordspacing

\bibitem{GHOST}
\BIBentryALTinterwordspacing
Y.~Sompolinsky and A.~Zohar, ``Secure high-rate transaction processing in
  bitcoin,'' in \emph{Financial Cryptography and Data Security - 19th
  International Conference, {FC} 2015, San Juan, Puerto Rico, January 26-30,
  2015, Revised Selected Papers}, 2015, pp. 507--527. [Online]. Available:
  \url{https://doi.org/10.1007/978-3-662-47854-7\_32}
\BIBentrySTDinterwordspacing

\bibitem{HotStuff}
\BIBentryALTinterwordspacing
M.~Yin, D.~Malkhi, M.~K. Reiter, G.~Golan{-}Gueta, and I.~Abraham, ``Hotstuff:
  {BFT} consensus with linearity and responsiveness,'' in \emph{Proceedings of
  the 2019 {ACM} Symposium on Principles of Distributed Computing, {PODC} 2019,
  Toronto, ON, Canada, July 29 - August 2, 2019.}, 2019, pp. 347--356.
  [Online]. Available: \url{https://doi.org/10.1145/3293611.3331591}
\BIBentrySTDinterwordspacing

\bibitem{bagaria2019prism}
V.~Bagaria, S.~Kannan, D.~Tse, G.~Fanti, and P.~Viswanath, ``Prism:
  Deconstructing the blockchain to approach physical limits,'' in
  \emph{Proceedings of the 2019 ACM SIGSAC Conference on Computer and
  Communications Security}, 2019, pp. 585--602.

\bibitem{Nakamoto2008Bitcoin}
S.~Nakamoto, ``Bitcoin: A peer-to-peer electronic cash system,''
  \emph{Consulted}, 2008.

\bibitem{OmniLedger}
\BIBentryALTinterwordspacing
E.~Kokoris{-}Kogias, P.~Jovanovic, L.~Gasser, N.~Gailly, E.~Syta, and B.~Ford,
  ``Omniledger: {A} secure, scale-out, decentralized ledger via sharding,'' in
  \emph{2018 {IEEE} Symposium on Security and Privacy, {SP} 2018, Proceedings,
  21-23 May 2018, San Francisco, California, {USA}}, 2018, pp. 583--598.
  [Online]. Available: \url{https://doi.org/10.1109/SP.2018.000-5}
\BIBentrySTDinterwordspacing

\bibitem{OHIE}
\BIBentryALTinterwordspacing
H.~Yu, I.~Nikolic, R.~Hou, and P.~Saxena, ``{OHIE:} blockchain scaling made
  simple,'' \emph{CoRR}, vol. abs/1811.12628, 2018. [Online]. Available:
  \url{http://arxiv.org/abs/1811.12628}
\BIBentrySTDinterwordspacing

\bibitem{RapidChain}
\BIBentryALTinterwordspacing
M.~Zamani, M.~Movahedi, and M.~Raykova, ``Rapidchain: Scaling blockchain via
  full sharding,'' in \emph{Proceedings of the 2018 {ACM} {SIGSAC} Conference
  on Computer and Communications Security, {CCS} 2018, Toronto, ON, Canada,
  October 15-19, 2018}, 2018, pp. 931--948. [Online]. Available:
  \url{https://doi.org/10.1145/3243734.3243853}
\BIBentrySTDinterwordspacing

\bibitem{Sharding2016}
\BIBentryALTinterwordspacing
L.~Luu, V.~Narayanan, C.~Zheng, K.~Baweja, S.~Gilbert, and P.~Saxena, ``A
  secure sharding protocol for open blockchains,'' in \emph{Proceedings of the
  2016 {ACM} {SIGSAC} Conference on Computer and Communications Security,
  Vienna, Austria, October 24-28, 2016}, 2016, pp. 17--30. [Online]. Available:
  \url{https://doi.org/10.1145/2976749.2978389}
\BIBentrySTDinterwordspacing

\bibitem{Chainspace}
\BIBentryALTinterwordspacing
M.~Al{-}Bassam, A.~Sonnino, S.~Bano, D.~Hrycyszyn, and G.~Danezis,
  ``Chainspace: {A} sharded smart contracts platform,'' in \emph{25th Annual
  Network and Distributed System Security Symposium, {NDSS} 2018, San Diego,
  California, USA, February 18-21, 2018}, 2018. [Online]. Available:
  \url{http://wp.internetsociety.org/ndss/wp-content/uploads/sites/25/2018/02/ndss2018\_09-2\_Al-Bassam\_paper.pdf}
\BIBentrySTDinterwordspacing

\bibitem{IOTA}
\BIBentryALTinterwordspacing
S.~Popov, O.~Saa, and P.~Finardi, ``Equilibria in the tangle,'' \emph{CoRR},
  vol. abs/1712.05385, 2017. [Online]. Available:
  \url{http://arxiv.org/abs/1712.05385}
\BIBentrySTDinterwordspacing

\bibitem{conflux}
\BIBentryALTinterwordspacing
C.~Li, P.~Li, W.~Xu, F.~Long, and A.~C. Yao, ``Scaling nakamoto consensus to
  thousands of transactions per second,'' \emph{CoRR}, vol. abs/1805.03870,
  2018. [Online]. Available: \url{http://arxiv.org/abs/1805.03870}
\BIBentrySTDinterwordspacing

\bibitem{PHANTOM}
\BIBentryALTinterwordspacing
Y.~Sompolinsky and A.~Zohar, ``{PHANTOM:} {A} scalable blockdag protocol,''
  \emph{{IACR} Cryptology ePrint Archive}, vol. 2018, p. 104, 2018. [Online].
  Available: \url{http://eprint.iacr.org/2018/104}
\BIBentrySTDinterwordspacing

\bibitem{LN}
J.~Poon and T.~Dryja, ``The bitcoin lightning network: Scalable off-chain
  instant payments,''
  \url{https://lightning.network/lightning-network-paper.pdf}, 2016, january
  14, 2016.

\bibitem{Plasma}
J.~Poon and V.~Buterin, ``Plasma: Scalable autonomous smart contracts,''
  \url{https://plasma.io/plasma.pdf}, 2017, august 11, 2017.

\bibitem{OptimisticRollup}
O.~Rollup, ``Rollup protocol,''
  \url{https://community.optimism.io/docs/protocol/2-rollup-protocol}, 2023,
  october 4, 2023.

\bibitem{dilley2016strong}
J.~Dilley, A.~Poelstra, J.~Wilkins, M.~Piekarska, B.~Gorlick, and
  M.~Friedenbach, ``Strong federations: An interoperable blockchain solution to
  centralized third-party risks,'' \emph{arXiv preprint arXiv:1612.05491},
  2016.

\bibitem{wood2016polkadot}
G.~Wood, ``Polkadot: Vision for a heterogeneous multi-chain framework,''
  \emph{White paper}, vol.~21, no. 2327, p. 4662, 2016.

\bibitem{thomas2015protocol}
S.~Thomas and E.~Schwartz, ``A protocol for interledger payments,'' \emph{URL
  https://interledger. org/interledger. pdf}, 2015.

\bibitem{garoffolo2020zendoo}
A.~Garoffolo, D.~Kaidalov, and R.~Oliynykov, ``Zendoo: A zk-snark verifiable
  cross-chain transfer protocol enabling decoupled and decentralized
  sidechains,'' in \emph{2020 IEEE 40th International Conference on Distributed
  Computing Systems (ICDCS)}.\hskip 1em plus 0.5em minus 0.4em\relax IEEE,
  2020, pp. 1257--1262.

\bibitem{sasson2014zerocash}
E.~B. Sasson, A.~Chiesa, C.~Garman, M.~Green, I.~Miers, E.~Tromer, and
  M.~Virza, ``Zerocash: Decentralized anonymous payments from bitcoin,'' in
  \emph{2014 IEEE symposium on security and privacy}.\hskip 1em plus 0.5em
  minus 0.4em\relax IEEE, 2014, pp. 459--474.

\bibitem{ben2014succinct}
E.~Ben-Sasson, A.~Chiesa, E.~Tromer, and M.~Virza, ``Succinct
  $\{$Non-Interactive$\}$ zero knowledge for a von neumann architecture,'' in
  \emph{23rd USENIX Security Symposium (USENIX Security 14)}, 2014, pp.
  781--796.

\bibitem{bowe2019recursive}
S.~Bowe, J.~Grigg, and D.~Hopwood, ``Recursive proof composition without a
  trusted setup,'' \emph{Cryptology ePrint Archive}, 2019.

\bibitem{gluchowski2019zk}
A.~Gluchowski, ``Zk rollup: scaling with zero-knowledge proofs,'' \emph{Matter
  Labs}, 2019.

\bibitem{hong2021pyramid}
Z.~Hong, S.~Guo, P.~Li, and W.~Chen, ``Pyramid: A layered sharding blockchain
  system,'' in \emph{IEEE INFOCOM 2021-IEEE Conference on Computer
  Communications}.\hskip 1em plus 0.5em minus 0.4em\relax IEEE, 2021, pp.
  1--10.

\bibitem{crain2021red}
T.~Crain, C.~Natoli, and V.~Gramoli, ``Red belly: A secure, fair and scalable
  open blockchain,'' in \emph{2021 IEEE Symposium on Security and Privacy
  (SP)}.\hskip 1em plus 0.5em minus 0.4em\relax IEEE, 2021, pp. 466--483.

\bibitem{silvano2020iota}
W.~F. Silvano and R.~Marcelino, ``Iota tangle: A cryptocurrency to communicate
  internet-of-things data,'' \emph{Future generation computer systems}, vol.
  112, pp. 307--319, 2020.

\bibitem{tairi20212}
E.~Tairi, P.~Moreno-Sanchez, and M.~Maffei, ``A2l: Anonymous atomic locks for
  scalability in payment channel hubs,'' in \emph{2021 IEEE Symposium on
  Security and Privacy (SP)}.\hskip 1em plus 0.5em minus 0.4em\relax IEEE,
  2021, pp. 1834--1851.

\bibitem{gavzi2019proof}
P.~Ga{\v{z}}i, A.~Kiayias, and D.~Zindros, ``Proof-of-stake sidechains,'' in
  \emph{2019 IEEE Symposium on Security and Privacy (SP)}.\hskip 1em plus 0.5em
  minus 0.4em\relax IEEE, 2019, pp. 139--156.

\bibitem{neu2021ebb}
J.~Neu, E.~N. Tas, and D.~Tse, ``Ebb-and-flow protocols: A resolution of the
  availability-finality dilemma,'' in \emph{2021 IEEE Symposium on Security and
  Privacy (SP)}.\hskip 1em plus 0.5em minus 0.4em\relax IEEE, 2021, pp.
  446--465.

\bibitem{tian2021enabling}
H.~Tian, K.~Xue, X.~Luo, S.~Li, J.~Xu, J.~Liu, J.~Zhao, and D.~S. Wei,
  ``Enabling cross-chain transactions: A decentralized cryptocurrency exchange
  protocol,'' \emph{IEEE Transactions on Information Forensics and Security},
  vol.~16, pp. 3928--3941, 2021.

\bibitem{sober2021voting}
M.~Sober, G.~Scaffino, C.~Spanring, and S.~Schulte, ``A voting-based blockchain
  interoperability oracle,'' in \emph{2021 IEEE International Conference on
  Blockchain (Blockchain)}.\hskip 1em plus 0.5em minus 0.4em\relax IEEE, 2021,
  pp. 160--169.

\bibitem{bunz2018bulletproofs}
B.~B{\"u}nz, J.~Bootle, D.~Boneh, A.~Poelstra, P.~Wuille, and G.~Maxwell,
  ``Bulletproofs: Short proofs for confidential transactions and more,'' in
  \emph{2018 IEEE symposium on security and privacy (SP)}.\hskip 1em plus 0.5em
  minus 0.4em\relax IEEE, 2018, pp. 315--334.

\bibitem{saleh2021blockchain}
F.~Saleh, ``Blockchain without waste: Proof-of-stake,'' \emph{The Review of
  financial studies}, vol.~34, no.~3, pp. 1156--1190, 2021.

\bibitem{grassi2021poseidon}
L.~Grassi, D.~Khovratovich, C.~Rechberger, A.~Roy, and M.~Schofnegger,
  ``Poseidon: A new hash function for $\{$Zero-Knowledge$\}$ proof systems,''
  in \emph{30th USENIX Security Symposium (USENIX Security 21)}, 2021, pp.
  519--535.

\bibitem{yang2020zero}
X.~Yang and W.~Li, ``A zero-knowledge-proof-based digital identity management
  scheme in blockchain,'' \emph{Computers \& Security}, vol.~99, p. 102050,
  2020.

\bibitem{LuuNZBGS16}
\BIBentryALTinterwordspacing
L.~Luu, V.~Narayanan, C.~Zheng, K.~Baweja, S.~Gilbert, and P.~Saxena, ``A
  secure sharding protocol for open blockchains,'' in \emph{Proceedings of the
  2016 {ACM} {SIGSAC} Conference on Computer and Communications Security,
  Vienna, Austria, October 24-28, 2016}, E.~R. Weippl, S.~Katzenbeisser,
  C.~Kruegel, A.~C. Myers, and S.~Halevi, Eds.\hskip 1em plus 0.5em minus
  0.4em\relax {ACM}, 2016, pp. 17--30. [Online]. Available:
  \url{https://doi.org/10.1145/2976749.2978389}
\BIBentrySTDinterwordspacing

\bibitem{GaiNB0W23}
\BIBentryALTinterwordspacing
F.~Gai, J.~Niu, I.~Beschastnikh, C.~Feng, and S.~Wang, ``Scaling blockchain
  consensus via a robust shared mempool,'' in \emph{39th {IEEE} International
  Conference on Data Engineering, {ICDE} 2023, Anaheim, CA, USA, April 3-7,
  2023}.\hskip 1em plus 0.5em minus 0.4em\relax {IEEE}, 2023, pp. 530--543.
  [Online]. Available: \url{https://doi.org/10.1109/ICDE55515.2023.00047}
\BIBentrySTDinterwordspacing

\end{thebibliography}

\end{document}